\documentclass[reprint,secnumarabic,amssymb, aps, prb]{revtex4-2}

\usepackage{graphicx}
\usepackage{graphics}
\usepackage[percent]{overpic}
\graphicspath{ {images/} }
\usepackage{amsmath}
\usepackage{mathtools}

\usepackage[utf8]{inputenc}
\usepackage{amsfonts}
\usepackage{amssymb}
\usepackage{float}
\usepackage{comment}
\usepackage{color,soul}
\usepackage{textcomp}
\usepackage{subcaption}
\usepackage{hyperref}
\usepackage[dvipsnames]{xcolor}
\usepackage{physics}
\begin{document}

%%%%%%%%%%%%%%%%%%%%%%%%%%%%%%%%%%%%%%%%%%%%%%%%%%%%%%%%%%%%%%%%%%%%%%%%%%%%%%%%%%%%%%%%%%%%%%%%%%%%%%%%
\title{Structural and Magnetic Properties of Co-rich bct Co$_{1-x}$Mn$_x$ Films}%

\author{Sean F Peterson}
\author{Yves U Idzerda}
%\email[REVTeX Support: ]{revtex@aps.org}
\affiliation{Department of Physics, Montana State University, Bozeman, Montana 59717, USA}
\date{\today}
\begin{abstract}
        \noindent{Thin-films of bct Co$_{1-x}$Mn$_x$ grown by molecular beam epitaxy on MgO(001) were measured to have an enhanced atomic magnetic moment of $2.52 \pm 0.07$ $\mu_\text{B}/\text{atom}$ beyond the pinnacle of the Slater-Pauling curve for Fe$_{1-x}$Co$_{x}$ with a moment of $2.42$ $\mu_\text{B}/\text{atom}$. The compositional variation of the average total moment for thin-film bct Co$_{1-x}$Mn$_x$ alloys is in stark contrast to the historical measurements of bulk fcc Co$_{1-x}$Mn$_x$. Generalized gradient approximation calculations reveal that significant improvements of this ferromagnetic forced bct phase on MgO(001) are possible via substrate selection. For example, bct Co$_{1-x}$Mn$_x$ films on MgO(001) are calculated to have lower atomic moments than those on substrates with smaller lattice constants such as GaAs(001), BaTiO$_3$(110), and SrTiO$_3$(110) which is predicted to increase the average atomic moment up to $2.61$ $\mu_\text{B}/\text{atom}$ and lead to increased structural stability and therefore thicker film growths.}
\end{abstract}
\maketitle
%\tableofcontents

%%%%%%%%%%%%%%%%%%%%%%%%%%%%%%%%%%%%%%%%%%%%%%%%%%%%%%%%%%%%%%%%%%%%%%%%%%%%%%%%%%%%%%%%%%%%%%%%%%%%%%%%
\section{Introduction}

The Slater-Pauling curve\cite{slater1936,pauling1938,williams1983} serves as a rough guide for the composition dependence of the average atomic moment of binary alloys, and achieves a maximum value for Fe$_{0.65}$Co$_{0.35}$ with a moment of $2.42$ $\mu_\text{B}/\text{atom}$. Historically, bulk Co$_{1-x}$Mn$_x$ alloys were observed to have a local face-centered cubic (fcc) structure and a rapidly decreasing average atomic moment with increasing Mn concentration\cite{cable1982,nakai1978}, resulting in a stark divergence from the Slater-Pauling rule. This is largely attributed to an antiferromagnetic (or ferrimagnetic) alignment between nearest neighbor Mn moments, resulting in a near-zero average Mn moment and a rapidly diminishing Co moment. 

However in recent years, Co$_{1-x}$Mn$_x$ films have been grown on GaAs(001)\cite{wu2001} and MgO(001)\cite{snowCoMn,wang2021,kunimatsu2021} substrates, allowing them to take on the forced\cite{liu1993} body-centered tetragonal (bct) structural phase as opposed to the stable fcc phase\cite{chambers1987}. Interestingly, this forced structural phase of Co$_{1-x}$Mn$_x$ was also found to enhance the ferromagnetic alignment of the Mn atoms\cite{wang2021,kunimatsu2021} with atomic moments approaching $3$ $\mu_\text{B}$, and resulted in a composition-dependent average moment reminiscent of the Slater-Pauling curve\cite{snowCoMn}. 

Large Mn moments aren't unique to bct Co$_{1-x}$Mn$_x$ films, but were also observed in Fe$_x$Co$_y$Mn$_z$ films\cite{snow2018,kashyap2018} as well as in Ni-rich fcc Ni$_{1-x}$Mn$_x$ alloys\cite{cable1974}. This high Mn moment was, in part, responsible for the impressive average moment of $(3.3 \pm 0.3)$ $ \mu_\text{B}/\text{atom}$ in bct Fe$_{0.09}$Co$_{0.62}$Mn$_{0.29}$ films on MgO(001)\cite{snow2018}, a significant improvement over the pinnacle of the Slater-Pauling curve for binary alloys. In fact, a high-moment ferromagnetic Mn state was first predicted in pure body-centered-cubic (bcc) Mn with a large lattice constant of $\sim3.1$ \AA, referred to as the high-spin state\cite{fuster1988,fry1987}. 

As the atomic moment of a material directly impacts the spin-polarization, these enhanced moment Co$_{1-x}$Mn$_x$ films lend themselves to technological applications as components in magnetic tunnel junctions (MTJs)\cite{kunimatsu2019,suzuki2021}. Early applications of sputtered Co$_{3/4}$Mn$_{1/4}$/MgO/Co$_{3/4}$Mn$_{1/4}$ showed promising results with a tunneling magnetoresistance (TMR) effect of $250\%$ at room temperature and greater than $600\%$ at low-temperature\cite{kunimatsu2020,elphick2021}. This effect was further increased when alloyed with Fe to create Fe$_{0.17}$(Co$_{0.79}$Mn$_{0.21}$)$_{0.83}$ which had a TMR effect of $350\%$ at room temperature and above $1000\%$ at low-temperatures\cite{ichinose2023}. 

MTJs are essential spintronics components\cite{julliere1975,miyazaki1995,moodera1995} with applications spanning from hard drive read heads\cite{kent2015} to magnetoresistive random access memory (MRAM)\cite{jimmy2006}. In order to improve the memory density of MRAM devices, perpendicularly magnetized MTJ (pMTJ) bits are written with a spin-transfer torque (STT) writing mechanism\cite{bhatti2017,yamamoto2023}, which leads to a TMR effect in excess of $100\%$ and is scalable to lower dimensions more so than a magnetic writing mechanism. Understanding the structural and magnetic properties of both bct Co$_{1-x}$Mn$_{x}$ and Fe$_{x}$Co$_{y}$Mn$_{z}$ films are of great interest to the continued development of MTJs and MRAM devices. 

The organization of the paper is as follows. In Section~\ref{sec:exp}\ref{sec:growth} the growth by molecular beam epitaxy (MBE) and measurement of bct Co$_{1-x}$Mn$_x$ films on MgO(001) is described briefly, with additional details reported elsewhere\cite{snowCoMn}. Section~\ref{sec:exp}\ref{sec:exp_results} investigates the variation of the moments of these Co$_{1-x}$Mn$_x$ films as a function of Mn concentration. Section~\ref{sec:exp}\ref{sec:comparison} compares the concentration dependent bct Co$_{1-x}$Mn$_x$ average moments with average moment measurements of other binary alloys commonly associated with the Slater-Pauling curve. In Section~\ref{sec:theory}\ref{sec:structural} ab-inito calculations were performed for a range of Co$_{1-x}$Mn$_x$ bct materials in order to characterize the Bain deformation of these materials as the in-plane lattice constant is artificially modified, which could be experimentally achieved by growth on different substrates. Section~\ref{sec:theory}\ref{sec:atomMag} considers further ab-initio calculations to determine how the Bain deformation of Co$_{1-x}$Mn$_x$ affects the average atomic moments, which were used to normalize the atomic moment data previously presented in Section~\ref{sec:exp}\ref{sec:exp_results}. Section~\ref{sec:conclusion} contains a brief conclusion which discusses how the results of these ab-initio calculation could be used to improve the average moment of bct Co$_{1-x}$Mn$_x$ films by growth on different substrates aside from MgO and how this could improve Co$_{3/4}$Mn$_{1/4}$-based MTJs.

%%%%%%%%%%%%%%%%%%%%%%%%%%%%%%%%%%%%%%%%%%%%%%%%%%%%%%%%%%%%%%%%%%%%%%%%%%%%%%%%%%%%%%%%%%%
\section{\label{sec:exp} Experiment}

\subsection{\label{sec:growth}Growth and Measurement}

Molecular beam epitaxy (MBE) was used to produce single-crystal epitaxial ultrathin Co$_{1-x}$Mn$_x$ films in the manner described previously\cite{snowCoMn}. A polished MgO(001) substrate was chosen for comparison with previous work and for its expected bcc structure. The MgO lattice constant is $4.21$ \AA, with an O-O distance of $2.98$ \AA, which is $5.3\%$ larger than the bcc lattice constant of bcc Co and $4.2\%$ larger than the assumed lattice constant of bcc Co$_{1/2}$Mn$_{1/2}$ (data from Table~\ref{tab:substrateMismatch}). Lattice parameters for bcc Co$_{1-x}$Mn$_x$ alloy at $x = 0$, $1/4$, $1/2$, and $1$ are calculated to be $2.83$ \AA, $2.85$ \AA, $2.86$ \AA, and $2.89$ \AA \hspace{.01in}, respectively, indicating that it is feasible to grow the bcc structure for the entire Co$_{1-x}$Mn$_x$ alloy range with only a $3$-$5\%$ lattice mismatch on MgO(001).

Before MBE growth, substrates were cleaned, dried, mounted, and annealed at $800^\circ$C. A Co$_{1-x}$Mn$_x$ film grown directly on MgO led to oxidation and poor epitaxial growth, so a $2$ nm Fe buffer layer was used for stabilization. Films were capped with a $3$ nm Al layer to prevent oxidation. Later, film oxidation was prevented without the Fe buffer by changing the MgO substrate vendor and modifying the cleaning procedure (heating to $850^\circ$C for 15 minutes), similar to other recent studies\cite{kunimatsu2021,wang2021}. The Co and Mn moments were verified to be independent of the presence of the Fe buffer layer.

Single-crystal epitaxial growth in the bcc structure was monitored using RHEED. Films with Mn concentrations $\leq 0.6$ retained the RHEED pattern for $4$ nm thick growths, but above $0.6$, the pattern was only retained up to $\sim 2$ nm thick growths. RHEED confirmed the body-centered structure and displayed a $45^\circ$ rotation of the film crystalline axes compared to the substrate. For compositions above $x = 0.7$, RHEED showed disruptions in the bcc structure from the start of growth\cite{snowCoMn}.  

The composition and atomic magnetic moments of these films were determined by X-ray measurements performed at beamlines 4.0.2 and 6.3.1 of the Advanced Light Source (ALS) at Lawrence-Berkeley National Laboratory. The composition was determined by X-ray absorption spectroscopy (XAS) and the atomic moment was determined by X-ray magnetic circular dichroism (MCD), the details of which are reported elsewhere\cite{snowCoMn}.

\subsection{\label{sec:exp_results}Results}

The MCD measurements for Co and Mn were normalized to the calculated atomic moments for bct Co$_{1-x}$Mn$_x$ at in-plane lattice constants, $a^\text{sub}$, consistent with the lattice mismatch of these materials grown on an MgO(001) substrate relative to their calculated bcc lattice constants ($a^\text{sub} = 2.91$ \AA, see TABLE~\ref{tab:latticeConstants} and \ref{tab:substrateMismatch}). 

\begin{figure}
\includegraphics[width=8cm]{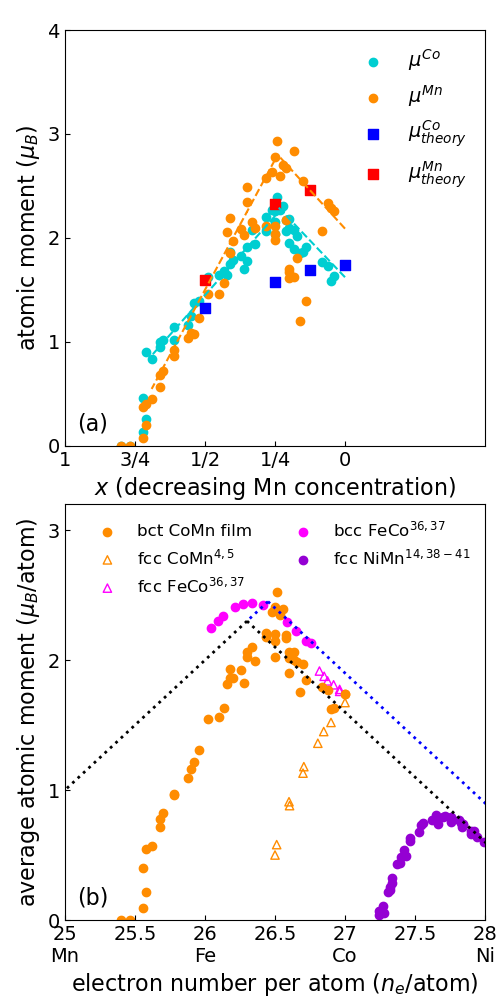}
\caption{\label{fig:atomicMag_exp} (a) MCD data of Co (cyan) and Mn (orange) measured for Co$_{1-x}$Mn$_{x}$ thin films grown on MgO(001) and normalized by calculated atomic magnetic moments. The Co atomic moment was normalized to $1.74$ $\mu_\text{B}$ for pure Co ($x = 0$) and the Mn atomic moment was normalized to minimize the residuals between the lines fit to the atomic Mn  moment (seen as the dashed orange lines) and the calculated atomic Mn moments $2.46$ $\mu_\text{B}$ for Co$_{7/8}$Mn$_{1/8}$, $2.33$ $\mu_\text{B}$ for Co$_{3/4}$Mn$_{1/4}$, and $1.60$ $\mu_\text{B}$ for Co$_{1/2}$Mn$_{1/2}$. Calculated Co and Mn atomic moments are represented by blue and red squares respectively. (b) The average atomic magnetic moment of bct Co$_{1-x}$Mn$_x$ on MgO (001) is shown in orange plotted against the electron number per atom alongside other transition metal alloys with Mn and the theoretical Slater-Pauling curve for late transition metals (black dotted line) and for Co-based alloys (blue dotted line) at electron numbers between that of Mn ($n_\text{e} = 25$) and Ni ($n_\text{e} = 28$).}
\end{figure}

The Co atomic moment was normalized to be $1.74$ $\mu_\text{B}$ for pure bct Co, ($x=0$), which is a well-known result for pure Co\cite{crangle1955,li1988,wu1992,bland1991}. The Mn atomic moment was normalized to minimize the residuals between the fit-lines for the atomic Mn saturation moment (see FIG.~\ref{fig:atomicMag_exp}(a) orange dashed lines) and three calculated magnetic moments at $x = (1/8, 1/4, 1/2)$ (see FIG.~\ref{fig:atomicMag_exp}(a) red squares). The Mn atomic moments were determined to be $\mu^\text{Mn} = 2.46$ $\mu_\text{B}$, $\mu^\text{Mn} = 2.33$ $\mu_\text{B}$, and $\mu^\text{Mn} = 1.60$ $\mu_\text{B}$ for bct Co$_{7/8}$Mn$_{1/8}$, bct Co$_{3/4}$Mn$_{1/4}$, and bct Co$_{1/2}$Mn$_{1/2}$, respectively.

The average atomic magnetic moment is the compositional average of the Co and Mn atomic magnetic moments, $\mu^\text{avg} = (1-x)\mu^\text{Co} + x\mu^\text{Mn}$. The average atomic moment can be plotted against the electron number per atom, $n_e$ (simply another way to represent $x$, the Mn concentration) in order to compare it to other magnetic materials from the Slater-Pauling curve, as seen in FIG.~\ref{fig:atomicMag_exp}(b). 

As the Mn concentration is increased from $x = 0$ to $1/4$ (electron number per atom decreases from $N_\text{e}/\text{atom} = 27$ to $26.5$), the average magnetic moment per atom increases linearly from $1.74$ $\mu_\text{B}/\text{atom}$ to $2.45$ $\mu_\text{B}/\text{atom}$ (see orange fit line in FIG.~\ref{fig:atomicMag_exp}(b)). The maximum average atomic moment occurs at $x = 0.24$ with a moment of $(2.52 \pm 0.07)$ $\mu_\text{B}/\text{atom}$. This increase in the average atomic moment is due to both an increase of the individual Co and Mn atomic moments and the increased Mn contribution to the average. 

The large variability of the experimental average atomic moment near $x = 1/4$ ($N_\text{e}/\text{atom} = 26.5$) reported in FIG.~\ref{fig:atomicMag_exp}(b) is due to the large variability in the average Mn moment reported in FIG.~\ref{fig:atomicMag_exp}(a). The Mn moments sample-to-sample variability arises from the distribution of equivalent compositional phases within the disordered alloy that have different average Mn moments, many of which are ferromagnetically aligned while others are ferrimagnetically aligned with adjacent Mn moments. 

For Co-rich alloys (Co$_{7/8}$Mn$_{1/8}$ and Co$_{3/4}$Mn$_{1/4}$) at the MgO(001) substrate in-plane lattice constant, for those Mn atoms that have a nearest-neighbors Mn atom, the two Mn moments are anti-parallel in their alignments, reducing the overall average Mn moment. Utilizing a random distribution of the complete set of possible atomic locations that properly account for the multiplicity of the redundant configurations, results in a significant reduced average Mn moment (see TABLE~\ref{tab:Co875Mn125_orders} and \ref{tab:Co75Mn25_orders}). While these configurations of Mn atoms within the Co lattice are possible (and even favored by the multiplicity of the state), they are energetically unfavorable states as they tend to be $\sim20$ eV above the most energetically favorable state for a given alloy whereas the other Mn configurations tend to be within $\sim5$ eV of the most favorable state.  These antiferromagnetic Mn alignment pairings are anticipated to appear in these MBE samples since the growth is performed in a non-thermodynamic manner resulting in Mn distributions that are subtly dependent on details of the growth and substrate conditions. 
 
 Previous calculations\cite{kunimatsu2021} of disordered Co$_{1-x}$Mn$_x$ alloys with 250 atom supercells with special quasi-random structures designed to mimic chemical disorder\cite{zunger1990} also found the presence of accidental, antiferromagnetic nearest-neighbor Mn pairs. However, far fewer of these antiferromagnetic pairs were reported than were observed experimentally and would be expected for purely random configurations of Co and Mn atoms in a 250 atom supercell. 
 
 The low Mn moment data points in FIG.~\ref{fig:atomicMag_exp}(a) when $x < 1/4$ result from anti-parallel Mn pairings and are reduced from the maxium attainable moment for that composition.  For comparison to the Slater-Pauling curve, these reduced moment values are excluded from the normalization fit procedure (the orange dashed line in FIG.~\ref{fig:atomicMag_exp}(a) only included the high-moment Mn data). 
 
In FIG.~\ref{fig:atomicMag_exp}(b) it can be seen that the average atomic moment of bct Co$_{1-x}$Mn$_x$ diverges from the typical Slater-Pauling curve (black dotted line), which assumes a fixed contribution of $\sim0.3$ $\mu_\text{B}$/atom from the electrons in the $sp$-band for the late transition metals\cite{moruzzi1978}. However, the deviation of the average moment of bct Co$_{1-x}$Mn$_x$ from the Slater-Pauling curve is consistent with the observation that Co alloys are magnetically strong and actually contribute $\sim0.45$ $\mu_\text{B}$/atom from the electrons in the $sp$-band\cite{williams1983}. This can be seen in FIG.~\ref{fig:atomicMag_exp}(b) where the bct Co$_{1-x}$Mn$_x$ average moments tend to be above the black dotted line and approach the blue dotted line near $n_e/\text{atom} = 26.5$

As the Mn concentration is increased further beyond $x=1/4$ to $\sim 0.7$ (electron number per atom decreases from $n_\text{e}/\text{atom}=26.5$ to $\sim25.6$), the average magnetic moment per atom decreases from $2.45\mu_\text{B}/\text{atom}$ to $0.56\mu_\text{B}/\text{atom}$. This dramatic decrease in the average atomic moment is typically attributed to antiferromagnetically aligned Mn atoms clustering together at these higher Mn concentrations\cite{snowCoMn,kunimatsu2021}. While this may be the case for sputtered samples where Mn clustering in this manner would likely not be in a bcc phase, the calculations in this work found that nearest-neighbor Mn-Mn pairings in Co$_{1/2}$Mn$_{1/2}$ are ferromagnetic, albeit with weaker moments than when Mn atoms are more distant from each other. 

The nearest-neighbor Mn-Mn pairings also becomes a more energetically favorable state in Co$_{1/2}$Mn$_{1/2}$ (see TABLE~\ref{tab:Co50Mn50_orders}). This reduction in the Mn moments is consistent with previous calculations on pure Mn, which notably predicted the existence of a high-spin and low-spin ferromagnetic state in bcc Mn with the transition occurring around $3.1$ \AA, consistent with Mn nearest-neighbor distances that resulted in a lower calculated atomic moment\cite{fry1987,fuster1988}. Additionally, the absence of an antiferromagnetic Co$_{1/2}$Mn$_{1/2}$ explains the lower variability in the Mn atomic moment data near $x = 1/2$ in FIG.~\ref{fig:atomicMag_exp}(a). 

Above a Mn concentration of $x=0.7$ (below $n_e/\text{atom}=25.6$ and where Mn atoms significantly outnumber Co atoms), the MCD signal for both Co and Mn, and thus the average atomic moment, is zero due to the collapse of the bct lattice structure as evidenced by a disruption to the RHEED pattern from the initial film growth\cite{snowCoMn}. A similar reduction in moment was measured in bct Fe$_{1-x}$Mn$_x$ thin films on MgO(001)\cite{idzerda2015}. This structural collapse is associated with an energetically favorable antiferromagnetic phase of Mn, which is the cause of the collapse of the average atomic moment. Other studies of bct Co$_{1-x}$Mn$_x$ alloys had a similar structural collapse with an increase in Mn concentration, however those studies' samples collapsed between $x = 0.34$ and $x = 0.5$\cite{kunimatsu2021}. This early collapse is common for sputtered films in comparison with films grown by MBE, resulting in the clustering of Mn atoms leading to the structural collapse of the bct lattice of the film. 

\subsection{\label{sec:comparison} Comparison to Previous Results}

In order to highlight the importance of the large average atomic moment of $(2.52 \pm 0.07)$ $\mu_\text{B}/\text{atom}$ observed in Co$_{0.76}$Mn$_{0.24}$ films on MgO(001), the compositional dependence of the average atomic moment is showcased alongside previous measurements on similar selected materials which contribute to the Slater-Pauling curve\cite{slater1936,pauling1938} in FIG.~\ref{fig:atomicMag_exp}(b). These previous measurements include: fcc Co$_{1-x}$Mn$_x$\cite{cable1982,nakai1978} to show how different this material can behave depending on structure, Fe$_{1-x}$Co$_x$\cite{williams1983,chikazumi1964,landolt1962} to show that the average atomic moment of  Co$_{0.76}$Mn$_{0.24}$ exceeds the maximum of the Slater-Pauling curve for binary alloys, and fcc Ni$_{1-x}$Mn$_{x}$\cite{cable1974,tange1978,kaya1931,sadron1932,piercy1953} to show that the enhanced average moment of a binary alloy with Mn isn't exclusive to Co$_{1-x}$Mn$_x$.

\subsubsection{fcc Co$_{1-x}$Mn$_x$}

Historic bulk measurements\cite{cable1982,nakai1978} of the average atomic moments of Co$_{1-x}$Mn$_x$ (see orange triangles in FIG.~\ref{fig:atomicMag_exp}(b)) lie in stark contrast to more recent thin film measurements\cite{snowCoMn,kunimatsu2021,wang2021}. This is mainly due to a structural difference between the historic bulk fcc samples and the thin film samples, which are bct forced structures. The thin-film bct Co$_{1-x}$Mn$_x$ samples were strongly ferromagnetic over a large Mn concentration, as has been shown in this work. Whereas, the fcc phase of Co$_{1-x}$Mn$_x$ often resulted in either an antiferromagnetic or ferrimagnetic alignment between Mn atoms. 

There is good agreement that a rapidly decreasing average Co$_{1-x}$Mn$_x$ atomic moment with increasing Mn concentration is due to the reversal of Mn moments originally aligned with the ferromagnetic Co moment to being anti-aligned. However, there is disagreement on whether the average Mn moment is aligned\cite{nakai1978} or anti-aligned\cite{cable1982} with the ferromagnetic Co moment. Cable et al. measured an average Mn moment anti-parallel to the relatively stronger Co moment ($\mu^\text{Co} = 0.80$ $\mu_\text{B}$) of $\mu^\text{Mn} = -0.11$ $\mu_\text{B}$ at a Mn concentration of $x = 0.244$. Nakai measured an average Mn moment parallel to the relatively weaker Co moment ($\mu^\text{Co} = 0.60$ $\mu_\text{B}$) of $\mu^\text{Mn} = 0.21$ $\mu_\text{B}$ at a Mn concentration of $x = 0.25$. 

Calculations performed in this work for fcc ($c/a^\text{sub} = \sqrt{2}$) Co$_{3/4}$Mn$_{1/4}$ support a ferrimagnetic Mn moment where the Mn atomic moment aligned to the ferromagnetic Co moments are magnetically weaker than the Mn atomic moment anti-aligned to them (further explained in Section~\ref{sec:theory}\ref{sec:Co75Mn25moment}). It should be noted that for one structural ordering of Co$_{3/4}$Mn$_{1/4}$ investigated in this work, it was energetically favorable for the Co moments to collapse to zero, while the Mn moments were antiferromagnetic.  See Appendix A for further details.

\subsubsection{Fe$_{1-x}$Co$_x$}

Alloys of Fe and Co are often seen as the pinnacle of the Slater-Pauling curve\cite{chikazumi1964,landolt1962,williams1983} (see pink circles in FIG.~\ref{fig:atomicMag_exp}(b)). This is a rather intuitive result, given that Fe and Co are the strongest elemental ferromagnets with atomic moments of $2.2$ $\mu_\text{B}$ and $1.7$ $\mu_\text{B}$ respectively. The largest average atomic moment occurs for the Fe$_{0.65}$Co$_{0.35}$ ($n_e/\text{atom} = 26.35$) alloy with a moment of $2.42$ $\mu_\text{B}/\text{atom}$. Surprisingly, this value can be contested by bct Co$_{1-x}$Mn$_x$ films on MgO(001), due to the impressively high atomic moments of Mn when not in the vicinity of other Mn atoms\cite{kashyap2018}. The largest average atomic moment occurs for the Co$_{0.24}$Mn$_{0.76}$ ($n_e/\text{atom} = 26.52$) alloy with a moment of $(2.52 \pm 0.07)$ $\mu_\text{B}/\text{atom}$ (the uncertainty is associated with the Mn moment normalization procedure). However, the calculations performed in this work suggest that both the atomic moments of Co and Mn may in fact be near their lowest for the ferromagnetic state. Growing these same bct Co$_{1-x}$Mn$_x$ films on GaAs(001) or BaTiO$_3$(110) substrates could result in Mn moments increasing by $\sim0.20$ - $0.35$ $\mu_\text{B}$ (average moments increasing to $2.57 - 2.61$ $\mu_\text{B}$ respectively, for $x = 1/4$).

A notable difference between the compositional dependence of the average atomic moments of bcc Fe$_{1-x}$Co$_x$ and both bct Co$_{1-x}$Mn$_x$ and Ni$_{1-x}$Mn$_x$ is that Fe$_{1-x}$Co$_x$ has a significantly broader peak near its maximum average moment. The rapid decline in the average moment with increasing Mn content in both bct Co$_{1-x}$Mn$_x$ and Ni$_{1-x}$Mn$_x$ is the result of Mn atoms occupying a low-spin state when in close-proximity to each other, a phenomenon first realized in calculations of bcc Mn\cite{fuster1988,fry1987}. However, as the Co content is increased in bcc Fe$_{1-x}$Co$_x$ beyond $x = 0.35$, the Co moments are unlikely to decrease regardless of their proximity to each other, as it's known that the strong magnetism of Co is relatively insensitive to structure\cite{liu1993}.

\subsubsection{fcc Ni$_{1-x}$Mn$_x$}

Similar to the bct Co$_{1-x}$Mn$_x$ films investigated in this work, fcc Ni$_{1-x}$Mn$_x$ alloys resulted in an initial enhancement of the average atomic moment before rapidly falling to a net-zero average moment with increasing Mn concentration (see purple circles in FIG.~\ref{fig:atomicMag_exp}(b))\cite{cable1974,tange1978,kaya1931,sadron1932,piercy1953}. The initial rise in the average atomic moment of these alloys can be attributed to the increasing Mn concentration with a large atomic moment of $3.50$ $\mu_\text{B}$ when $x = 0.05$, whereas the Ni atomic moment is a mere $0.58$ $\mu_\text{B}$\cite{cable1974}. As the Mn concentration is increased, the Mn moment decreases rapidly and the Ni moment decreases gradually beyond $x = 0.10$, which is why the average atomic moment peaks around $n_e/\text{atom} = 27.7$. 

While this rapid decrease in the average Mn moment is often attributed to antiferromagnetic nearest-neighbor Mn pairs, it seems unlikely that such a mechanism would lead to a collapse of the average Mn moment at such a small Mn concentration of $x < 1/4$. An atomistic model of the atomic moments\cite{kouvel1969} predicts that Mn atoms with three or more Mn nearest-neighbors would have their spin-reversed\cite{tange1978}, which is an unlikely ordering to appear in the structure of Ni$_{3/4}$Mn$_{1/4}$ for a significant portion of the Mn atoms. 

Calculations of fcc Ni$_{1-x}$Mn$_x$ at Mn concentrations of $x = 1/8$ and $1/4$ reveal that the Mn moments are ferromagnetically aligned to the Ni atoms. At low Mn concentrations of $x = 1/8$, the Mn atoms were found to occupy a high-spin state with a moment of $\sim 3$ $\mu_\text{B}$ to be aligned with the Ni moments of $\sim 0.6$ $\mu_\text{B}$ (see TABLE \ref{tab:NiMn} in Appendix \ref{sec:NiMn}). This is consistent with measurements of low Mn-concentration alloys, which were observed to have an average Mn and Ni moment of $(3.18 \pm 0.07)$ $\mu_\text{B}$ and $0.53$ $\mu_\text{B}$ respectively\cite{cable1974}. As the Mn concentration was increased to $x = 1/4$, both the Mn and Ni atomic moments decreased. However, some orderings retained a high-spin state with Mn moments ranging from $2.5$ - $3.0$ $\mu_\text{B}$ while others transitioned to a low-spin state with a Mn moment of $1.7$ - $2.0$ $\mu_\text{B}$. Cable et. al.\cite{cable1974} observed a decreasing average atomic moment of fcc Ni$_{1-x}$Mn$_x$ alloys with increasing Mn concentrations above $x = 1/8$ due to a decrease in both the average Mn and Ni atomic moments, this is consistent with the Mn atoms occupying a low-spin state. The reduction in the average Mn atomic moment with increasing Mn concentration is due to the existence of a low-spin ferromagnetic Mn state rather than a ferrimagnetic Mn state. This is similar to what was observed in the Mn moments of bct Co$_{1-x}$Mn$_x$ films, where a reduced Mn moment aligned parallel to the Co or Ni moments resulted in a rapid decrease in the average atomic moment with increasing Mn concentration.

\section{\label{sec:theory} Theoretical Calculations}

\subsection{\label{sec:structural} Structural Calculations}

To fully understand these experimental measurements, structural calculations for the Co-rich phases of bct Co$_{1-x}$Mn$_x$ alloys when $x = (0, 1/4, 1/2)$ and bct Mn when $x = 1$ were performed using a Projector Augmented plane-Wave (PAW) basis with the Perdew-Burke-Ernzerhof (PBE) Generalized Gradient Approximation (GGA) exchange-correlation functional with a $820$ eV kinetic energy cutoff for the plane-wave basis and a $6 \times 6 \times 6$ $\mathbf{k}$-grid per unit cell (Co$_{3/4}$Mn$_{1/4}$ had a smaller $\mathbf{k}$-grid of $6 \times 6 \times 3$ to match the symmetry of its supercell in $\mathbf{k}$-space), which resulted in the convergence of lattice constants to within $\sim 50$ $\mu$\AA. The cells used to calculate the structural distortions were a single bct unit cell for Co, Co$_{1/2}$Mn$_{1/2}$, and Mn. The Co$_{3/4}$Mn$_{1/4}$ calculations required a minimum of 2 bct unit cells to achieve the proper ratio of Co and Mn atoms in the supercell. For each alloy composition, a variety of supercells were used for the structural relaxation calculation. All yielded results within $\sim5$ m\AA \hspace{0.01in} of each other regardless of the atomic configurations. 

To implement this procedure, the Co$_{1-x}$Mn$_x$ bct lattices were determined by first identifying the undistorted bcc lattice constant for the alloy, $a^\text{bcc}$, which are shown in TABLE~\ref{tab:latticeConstants}, then varying the in-plane lattice constant by $~15\%$ above and below the $a^\text{bcc}$ value at intervals of $0.01$ \AA. In ultrathin film growth epitaxy, an overlayer's in-plane lattice constant, $a$, is typically the same as the substrate's surface in-plane lattice constant, $a^\text{sub}$, (perhaps with a rotation of the overlayer's unit cell relative to the substrate). This is due to the relative strength of the interfacial strain present in thin-films. The strain created within the film by this forced in-plane epitaxy is accommodated by a relaxation of the out-of-plane overlayer lattice constant, $c$. By minimizing the total energy of the film, this distortion can be determined. 

It is informative to report these results in terms of the $c(a)/a$ ratio (where it is explicitly noted that this ratio is a function of the in-plane lattice constant). Further, because of the forced epitaxy, $c(a)/a$ is replaced with $c(a^\text{sub})/a^{\text{sub}}$ to more clearly note that the overlayer distortion is due to the effect of the substrate. Since the bcc and fcc structures are both special cases of the body-centered lattice group with $c/a^\text{sub} = 1$ or $c/a^\text{sub} = \sqrt{2}$, respectively, the electronic and magnetic properties of both structures can be determined by considering the general case of the bct system while simultaneously allowing for easier comparison of these calculations with literature values for the bcc and fcc high-symmetry structures.

\begin{table}[h!]
\caption{\label{tab:latticeConstants} Comparison between calculated lattice parameters for bcc, $a^\text{bcc}$, and fcc, $a^\text{fcc}$, Co$_{1-x}$Mn$_x$ and their experimentally measured/extrapolated counterparts. The body-centered-cubic (bcc) and face-centered-cubic (fcc) structures are special cases of the body-centered lattice group with $c/a^\text{sub} = 1$ or $\sqrt{2}$ respectively, therefore the bcc lattice constant is defined as $a^\text{bcc} = a^\text{sub}$ when $c/a^\text{sub} = 1$ and the fcc lattice constant is defined as $a^\text{fcc} = a^\text{sub}\sqrt{2}$ when $c/a^\text{sub} = \sqrt{2}$. These special cases can be seen in FIG.~\ref{fig:c_over_a_vs_a} when the $c(a^\text{sub})/a^\text{sub}$ curves cross the black dotted lines at $1$ and $\sqrt{2}$.}
\begin{ruledtabular}
\begin{tabular}{c|cccc}
$x$ & $0$ & $1/4$ & $1/2$ & $1$\\
\hline
$a^\text{bcc}$ & $2.767$ \AA & $2.773$ \AA & $2.786$ \AA & $2.792$ \AA\\
$a^\text{fcc}$ & $3.458$ \AA & $3.474$ \AA & $3.487$ \AA & $3.516$ \AA\\
\hline
$a^\text{bcc}_\text{exp}$ & $2.83$ \AA\cite{prinz1985} & $2.845$ \AA\footnote{\label{linearNote}Linearly extrapolated between Co and Mn lattice constants} & $2.86$ \AA$^\text{a}$ & $2.89$ \AA\cite{jonker}\footnote{Calculated from the assumption bcc Mn has the same atomic density as $\alpha$-Mn.}\\
$a^\text{fcc}_\text{exp}$ & $3.56$ \AA\cite{chambers1987} & $3.575$ \AA$^\text{a}$ & $3.59$ \AA$^\text{a}$ & $3.62$ \AA\cite{jin1994}\\
\end{tabular}
\end{ruledtabular}
\end{table}

In TABLE~\ref{tab:latticeConstants} it can be seen that the relaxed bcc lattice constants for Co$_{1-x}$Mn$_x$ calculated in this work are roughly $3\%$ lower than have been experimentally measured\cite{prinz1985} for bcc Co or calculated for bcc Co$_{3/4}$Mn$_{1/4}$ and bcc Co$_{1/2}$Mn$_{1/2}$ from the assumption that bcc lattice constant would vary linearly between the bcc Co lattice constant and the bcc Mn lattice constant. The fcc lattice constants were also calculated in this work by allowing the crystals to undergo a Bain deformation\cite{bain} from a bct structure to a fcc structure when $c/a^\text{sub} = \sqrt{2}$\cite{dmitriev1991}. The calculated fcc lattice constants were also found to be roughly $3\%$ lower than those measured experimentally for Co\cite{chambers1987} and the linearly extrapolated Co$_{3/4}$Mn$_{1/4}$ and Co$_{1/2}$Mn$_{1/2}$ values from the measured fcc Mn lattice constant\cite{jin1994}.

\begin{figure}[t!]
\begin{centering}
\includegraphics[width=8cm]{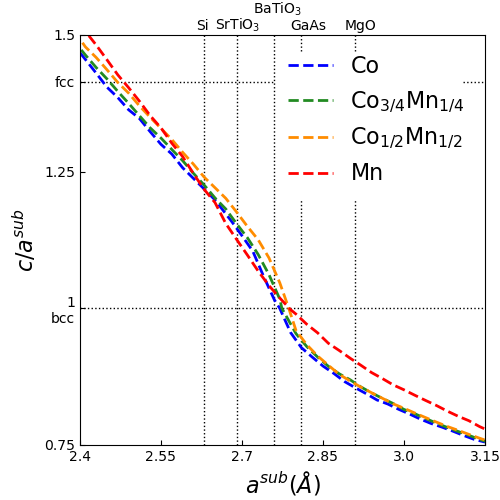}
\caption{Ratio of out-of-plane lattice constant, $c$, to epitaxially enforced in-plane lattice constant, $a^\text{sub}$, as the in-plane lattice constant is varied. Horizontal black dotted lines are included to show when $c/a^\text{sub} = 1$, corresponding to a bcc structure, and when $c/a^\text{sub} = \sqrt{2}$, corresponding to an fcc structure appearing due to the Bain deformation of a bct lattice. Vertical black dotted lines represent lattice constants of common Co substrates near its bcc lattice constant.} 
\label{fig:c_over_a_vs_a}
\end{centering}
\end{figure}

It can be seen in FIG.~\ref{fig:c_over_a_vs_a} that $c(a)/a^{\text{sub}}$ results are essentially monotonic and linear over the regions associated with the fcc and bcc structures. The inverse of the slopes in those linear regions are related to the Poisson ratio and are indicative of which high symmetry base structural phase the distortion is derived from. The Co, Co$_{3/4}$Mn$_{1/4}$, and Co$_{1/2}$Mn$_{1/2}$ all appear to have three distinct structural phases, whereas Mn only appears to have two.

The more common substrates for Co deposition have been organized in TABLE~\ref{tab:substrateMismatch} along with the corresponding in-plane lattice constants associated with them. Additionally, the fractional amount that they modify the known lattice constants of Co$_{1-x}$Mn$_x$ have also been tabulated. Since the calculated bcc lattice constants for these materials are not exactly equivalent to the known experimental values, it was more appropriate to consider a fractional difference between $a^\text{bcc}$ and $a^\text{sub}$ as opposed to using the experimental in-plane lattice constant for a given substrate. These scaled substrate values, specifically for Co$_{3/4}$Mn$_{1/4}$ although they aren't very different for the Co-rich Co$_{1-x}$Mn$_x$ alloys, were included in FIG.~\ref{fig:c_over_a_vs_a} to demonstrate the degree of tetragonal distortion each of these substrates created. The materials investigated in Section~\ref{sec:exp} were all deposited on MgO(001) substrates, and as such were calculated with a $\sim5\%$ larger in-plane lattice constant than the bcc lattice constant, which corresponds to $a^\text{sub}=2.91$ \AA.

\begin{table}[h!]
\caption{\label{tab:substrateMismatch} Modification of the bct Co$_{1-x}$Mn$_x$ in-plane lattice constant relative to the experimental bcc lattice constant for a selection of common substrates.}
\begin{ruledtabular}
\begin{tabular}{c|cccc}
Substrate & $a^\text{sub}_\text{exp}$ & $\frac{a^\text{sub}_\text{exp}}{a^\text{bcc}_\text{exp}(x=0)}$ & $\frac{a^\text{sub}_\text{exp}}{a^\text{bcc}_\text{exp}(x=\frac{1}{4})}$ & $\frac{a^\text{sub}_\text{exp}}{a^\text{bcc}_\text{exp}(x=\frac{1}{2})}$\\
\hline
fcc Si(110) & $2.70$ \AA & $0.954$ & $0.949$ & $0.944$\\
SrTiO$_3$(110) & $2.76$ \AA & $0.975$ & $0.970$ & $0.965$\\
BaTiO$_3$(110) & $2.83$ \AA & $1.0$ & $0.995$ & $0.990$\\
GaAs(001) & $2.88$ \AA & $1.018$ & $1.012$ & $1.007$\\
MgO(001) & $2.98$ \AA & $1.053$ & $1.048$ & $1.042$\\
\end{tabular}
\end{ruledtabular}
\end{table}

\subsubsection{Structural evolution of Co-rich Co$_{1-x}$Mn$_x$}

\begin{figure}[t!]
\begin{centering}
\includegraphics[width=8cm]{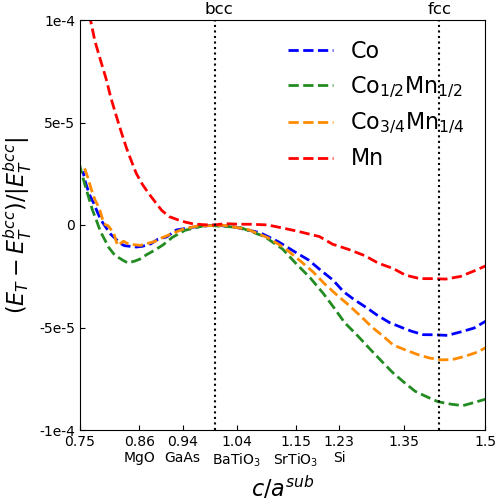}
\caption{Total energy difference, $E_T(c/a^\text{sub}) - E_T^\text{bcc}$, normalized by $|E_T^\text{bcc}|=|E_T(c/a^\text{sub}=1)|$ plotted as a function of $c/a^\text{sub}$ with black dotted guidelines at the bcc ($c/a^\text{sub}=1$) and fcc ($c/a^\text{sub}=\sqrt{2}$) phases. The tick-marks for $c/a^\text{sub}$ are labeled to represent the corresponding $c/a^\text{sub}$-ratio from common experimental substrates on Co$_{3/4}$Mn$_{1/4}$. This is done to demonstrate which structural phases of which Co$_{1-x}$Mn$_x$ alloys are stable and which are forced structures.} 
\label{fig:ET_vs_c_over_a}
\end{centering}
\end{figure}

When the Co-rich Co$_{1-x}$Mn$_x$ alloys are grown on substrates which force their in-plane lattice constants to go sufficiently below their bcc lattice constants ($\sim 5\%$ lower), the lattices can take on an fct structure as evidenced by the slope of $c(a^\text{sub})/a^\text{sub}$ in this region compared to the slope near $c/a^\text{sub} = \sqrt{2}$, of the fcc phase. Co has been experimentally shown to have a stable fcc phase\cite{chambers1987}, which is confirmed in this work by the existence of a stable total energy minimum when $c/a^\text{sub} = \sqrt{2}$, (see FIG.~\ref{fig:ET_vs_c_over_a}). Similarly, the Co$_{3/4}$Mn$_{1/4}$ and Co$_{1/2}$Mn$_{1/2}$ phases also have a minimum in their total energies near $c/a^\text{sub} = \sqrt{2}$, indicating that these materials have a stable fct phase as well.

As the substrate lattice constant approaches the bcc value, $0.95 a^\text{bcc} < a^\text{sub} < 1.05 a^\text{bcc}$, the Co-rich Co$_{1-x}$Mn$_x$ alloys undergo a Bain deformation and relax to a bct structure where the $c(a^\text{sub})/a^\text{sub}$ data in FIG.~\ref{fig:c_over_a_vs_a} for this region has a noticeably larger slope (and therefore a lower Poisson ratio). The bcc phase of the Co-rich Co$_{1-x}$Mn$_x$ alloys was confirmed to be a forced structure due to a local total energy maximum at the bcc phase when $c/a^\text{sub} = 1$, see FIG.~\ref{fig:ET_vs_c_over_a}. In earlier works, the bcc phase of Co was considered to be metastable\cite{idzerda1989,idzerda1990}, but more recent works have determined it to be a forced structure\cite{liu1993} which collapses to a hcp phase around a thickness of $50$\AA\cite{riedi1987}. Due to the absence of a stable bcc phase in bulk Co, these materials typically undergo a Martensitic transition directly from fcc to hcp\cite{menshikov1985,toledano2001}. However, due to the relatively high interfacial energies of epitaxial thin-films, the bcc Co$_{1-x}$Mn$_x$, thin-films can undergo a Bain deformation from fcc to bcc with a commensurate structural softening\cite{elphick2021}. This is also why bct Co$_{1-x}$Mn$_x$ has only been observed in thin films\cite{prinz1985,idzerda1990}, where the interfacial energies between the film and the substrate are larger than the bulk strain energies.

When the substrate lattice constant is sufficiently larger than the bcc lattice constant, $a^\text{sub} > 1.05 a^\text{bcc}$, the Co-rich Co$_{1-x}$Mn$_x$ alloys undergo another structural transition and the $c(a^\text{sub})/a^\text{sub}$ has a reduced slope in comparison to the bct region. This region is associated with the hcp phase, which is a stable phase of Co\cite{nishizawa}, and is associated with the local energy minimum at $c/a^\text{sub} = 0.83$ with a total energy $E_T(c/a^\text{sub} \approx 0.83)$ in FIG.~\ref{fig:ET_vs_c_over_a}. It should be noted that the lattice constants at which this hcp-like phase appears is close to those of the bct films grown on the MgO(001) substrate ($a^\text{sub} = 2.91$\AA, $c/a^\text{sub} = 0.86$ ratio) used in this work. It is likely  because of this proximity of the in-plane distortions caused by the MgO(001) substrate to this hcp-like structure that in thicker ($20$ nm) Co$_{1-x}$Mn$_x$ films on MgO(001) than were used in this work ($4$ nm) that the bcc phase appeared alongside a hcp phase\cite{wang2021}. In another study\cite{elphick2021}, it was found that the lattice constant of Co$_{1-x}$Mn$_x$ tended to increase by about $0.1$ \AA \hspace{0.01in} between the top and bottom layers of a MTJ stack. This is likely the result of the bct structure relaxing into a hcp phase.

This hcp-like structure can be compared to a true hcp phase by comparing the spacings between nearest-neighbor atoms in both cases. In the calculated hcp-like phase when $c/a^\text{sub} \approx 0.83$, the nearest-neighbor distance in the bct unit cell is along the $[111]$-axis with a value of $a^\text{hcp-like} = 2.44$ \AA. In a true hcp structure, the nearest-neighbor distance is between atoms which lie in the hexagonal plane separated by $a^\text{hcp}$. Experimentally, hcp Co has an in-plane lattice parameter\cite{arblaster2018} of $a^\text{hcp,Co} = 2.51$\AA. The calculated in-plane lattice constant of hcp-like Co is, similar to the bcc/fcc lattice parameters, roughly $3\%$ lower than the experimental in-plane lattice constant of hcp Co. However, properly modeling a bcc $\rightarrow$ hcp transition cannot be achieved through a Bain deformation alone as was considered in this work. Rather, it requires the Burgers mechanism\cite{burgers,dmitriev1991}. In bulk Co, the material would simply undergo a fcc $\rightarrow$ hcp (Martensitic) transition\cite{yang2018} since bcc Co is a forced structure and unattainable in bulk systems where the epitaxial strain of the substrate is negligible to the lattice.

\subsubsection{Structural evolution of Mn}

When the substrate lattice constant is less than the bcc lattice constant, $a^\text{sub} < 0.97 a^\text{bcc}$, Mn is in the fct phase. This can be seen in FIG.~\ref{fig:c_over_a_vs_a} because the slope in this region is roughly the same as that when $c/a^\text{sub}=\sqrt{2}$ (the fcc phase). The fcc phase of Mn is stable in this model, which can be seen from the local total energy minimum at $c/a^\text{sub} = \sqrt{2}$ in FIG.~\ref{fig:ET_vs_c_over_a}. This result is consistent with experiment, as Mn is known to have a stable fcc phase\cite{jin1994}.

When $a^\text{sub} > 0.97 a^\text{bcc}$, Mn goes through a structural transition to the bct phase as evidenced by the change in the Poisson ratio. Unlike Co in this model, the bcc Mn phase is in fact a stable phase (albeit weakly stable). This can be seen from the local total energy minimum when $c/a^\text{sub} = 1$, see FIG.~\ref{fig:ET_vs_c_over_a}. While some early works hint at the growth of bcc Mn at room temperature\cite{heinrich1987,jonker}, they actually demonstrate a bct phase with an in-plane lattice constant of $a = 2.89$ \AA \hspace{.01in} and $c/a = 1.15$, which can also be accurately described as an fct phase with $c/a = 0.81$\cite{egelhoff1990}. The true bcc phase of Mn, known as $\delta$-Mn, is a high-temperature phase which exists at temperatures well above any reasonable Curie or N\'{e}el\cite{hafner2003}. However, the $\alpha$-Mn phase  can be deposited on MgO(001) substrates\cite{socha2014} and is similar to a bcc phase, but requires 58 atoms per unit cell to model. Therefore, it is likely that this local minimum in the total energy is actually associated with the $\alpha$-Mn phase, which is outside ther scope of this work.

\subsection{\label{sec:atomMag}Atomic Moment Calculations}

Ising model atomic moment calculations, which restricted the moments to lie along the perpendicular $c$-axis, for bct Co$_{1-x}$Mn$_x$ alloys when $x = (0, 1/4, 1/2, 1)$ were performed for the $a^\text{sub}$-values in the previous section and their corresponding calculated $c/a^\text{sub}$-ratios shown in FIG. ~\ref{fig:c_over_a_vs_a}. A PAW basis was employed with the spin-resolved PBE GGA exchange-correlation functional with a $820$ eV kinetic energy cutoff for the plane-wave basis and a $10 \times 10 \times 10$ $\mathbf{k}$-grid per unit cell. Co$_{3/4}$Mn$_{1/4}$ and Co$_{1/2}$Mn$_{1/2}$ had reduced $\mathbf{k}$-grids to match the symmetry of their supercells in $\mathbf{k}$-space, which resulted in atomic moments of the bct materials to be converged within $\sim 0.005$ $\mu_\text{B}$. The supercells for each alloy consist of the minimum number of bct unit cells to allow for calculations with two Mn atoms in the supercell.  Mn atoms tend to prefer to align anti-parallel to each other in fcc calculations when two Mn atoms were considered in the supercell\cite{wu2001}. Additionally, two separate supercells were considered when the supercell for an alloy wasn't cubic, this was to allow for all possible configurations of the supercell relative to the unique $c$-axis of the Bain deformation. Every unique combination of Mn positions within these supercells was considered and averaged together, weighted by the multiplicity of the state, to approximate a disordered alloy. See Appendix~\ref{sec:superlattices} for the details of these calculations. 

\begin{figure}[t!]
\begin{centering}
\includegraphics[width=8cm]{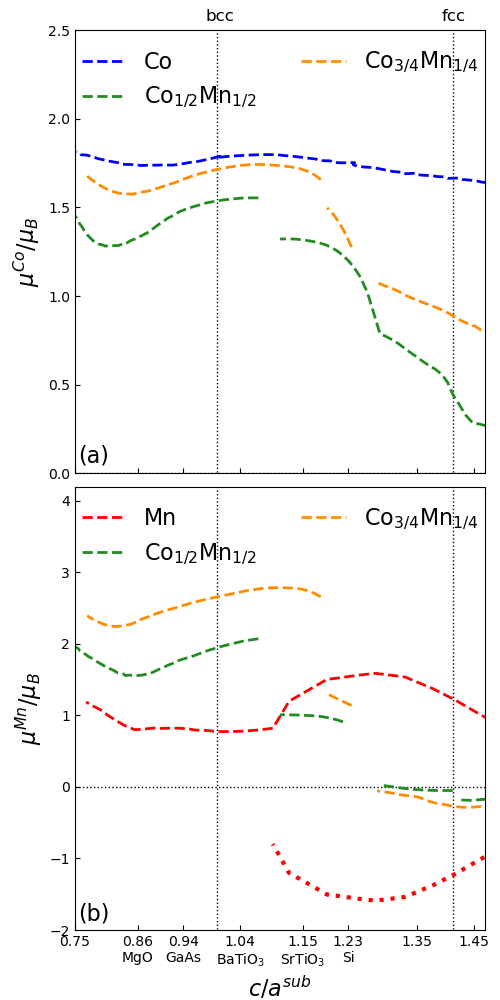}
\caption{Atomic magnetic moments in units of the Bohr magneton ($\mu_\text{B}$) as the Co$_{1-x}$Mn$_x$ alloys undergo a Bain deformation from bcc to fcc. (a) average atomic Co moment of pure bct Co (dashed blue), bct Co$_{3/4}$Mn$_{1/4}$ (dashed orange), and bct Co$_{1/2}$Mn$_{1/2}$ (dashed green). (b) average atomic Mn moment of bct Co$_{3/4}$Mn$_{1/4}$ (dashed orange) and bct Co$_{1/2}$Mn$_{1/2}$ (dashed green) and atomic Mn moments of pure bct Mn (dashed/dotted red) where dotted lines show where the two Mn moments are inequivalent in the antiferromagnetic state. The tick-marks for $c/a^\text{sub}$ are labeled to represent the corresponding $c/a^\text{sub}$-ratio from common experimental substrates on Co$_{3/4}$Mn$_{1/4}$.} 
\label{fig::atomicMag}
\end{centering}
\end{figure}

\subsubsection{Co moment}

For pure Co, the system remained ferromagnetic regardless of the structure (whether it be bct or fct) with a nearly constant atomic moment of $\sim1.7$ $\mu_\text{B}$. This is expected because the strong magnetism of Co is relatively insensitive to structure\cite{liu1993}. In general, the Co moment was slightly enhanced when the in-plane lattice constant was increased and the out-of-plane lattice constant was allowed to decrease with an atomic Co moment of $\sim 1.8$ $\mu_\text{B}$. The Co moment was slightly reduced when the in-plane lattice constant was decreased and the out-of-plane lattice constant increased resulting in an atomic Co moment of $\sim 1.6$ $\mu_\text{B}$ beyond the bcc $\rightarrow$ fcc transition. 

Early experimental works determined the atomic moment of bcc Co to be around $1.4\mu_\text{B}.$\cite{prinz1985} However, this relatively low atomic moment was later found to be the result of As intermixing near the interface with the substrate, and the pure Co moment at the center of the sample was measured to be $1.7$ $\mu_\text{B}.$\cite{bland1991} This is in good agreement with the bcc Co moment calculated in this work of $1.74$ $\mu_\text{B}$. The fcc phase of Co was previously measured\cite{crangle1955,li1988,wu1992} to be $1.62$ $\mu_\text{B}$, which is also in good agreement with the fcc Co moment of $1.66$ $\mu_\text{B}$. These values were also in good agreement with previous calculations\cite{liu1993,wu2001}.

\subsubsection{\label{sec:Co75Mn25moment}Co$_{3/4}$Mn$_{1/4}$ moment}

A disordered alloy of Co$_{3/4}$Mn$_{1/4}$ was approximated by considering seven distinct ordered alloys each with four bct unit cells per supercell. These ordered alloys considered every unique arrangement of the Mn atoms and consisted of a square of bct cells with both axes in an $a$-direction ($a$-class, see FIG.~\ref{fig:supercells_Co75Mn25} (a), (c), (e), and (g)), and with one axis along the $c$-direction and the other along an $a$-axis ($c$-class, see FIG.~\ref{fig:supercells_Co75Mn25} (b), (d), (f), and (h)). This was done in order to calculate magnetic states with two unique Mn atoms which could align anti-parallel to each other, a solution not considered in previous calculations\cite{wu2001} for this Mn concentration which found a single Mn moment anti-aligned with the Co moments. Of the seven ordered alloys considered in this work for $x = 1/4$, only five were averaged to model the disordered alloy because the two ordered alloys where the Mn atoms were nearest-neighbors resulted in an anti-parallel alignment of the Mn moments in the bcc phase, an energetically unfavorable configuration by about $20$ eV over the most favorable magnetic state (see TABLE~\ref{tab:Co75Mn25_orders}). 

The exclusion of these anti-parallel Mn moment states is also supported by previous calculations\cite{kunimatsu2021}, which discovered anti-parallel Mn moments due to accidental Mn nearest-neighbor pairs, but at a rate far lower than would be expected of purely randomly placed Mn atoms. In that work, SQSs were generated with 250 atoms to mimic chemical disorder and approximate the ensemble average\cite{zunger1990} with a single supercell, so the rarity of the anti-parallel Mn moment state is a testament to its energetic unfavorability. The moments of these alloys were calculated as a function of $c/a^\text{sub}$ and weighted by their multiplicities to approximate a disordered alloy. See Appendix \ref{sec:superlattices}\ref{sec:Co75Mn25} to see this in greater detail.

When $c/a^\text{sub} < 1.18$, all the Co and Mn moments are ferromagnetically aligned with an average Co moment ranging from $1.58$ $\mu_\text{B}$ to $1.74$ $\mu_\text{B}$ and an average Mn moment ranging from $2.25$ $\mu_\text{B}$ to $2.79$ $\mu_\text{B}$. The weakest ferromagnetic state occurs when $c/a^\text{sub} = 0.82$, near the hcp-like state discussed previously in Section \ref{sec:theory}\ref{sec:structural} and the tetragonal distortion associated with Co$_{3/4}$Mn$_{1/4}$ on MgO. While the strongest ferromagnetic state occurs when $c/a^\text{sub} = 1.09$, near the tetragonal distortion associated with growth on BTO or STO substrates. 

Above $c/a^\text{sub} = 1.18$ and below $1.23$, the average Mn moment abruptly reduces to $\sim 1.2$ $\mu_\text{B}$. This is due to regions of the disordered Co$_{3/4}$Mn$_{1/4}$ alloy with in-plane Mn-Mn next-nearest-neighbor pairs separated by Co atoms suddenly reversing the direction of one of the Mn moments to being ferrimagnetically aligned with the other, where the weaker Mn moment remains aligned with the Co moments. This ferrimagnetic transition in the Mn moments also reduces nearby Co moments abruptly by $\sim 0.2$ $\mu_\text{B}$. Simultaneously, other regions of the disordered alloy with Mn-Mn next-nearest-neighbor pairs separated by a Co atom exclusively in the out-of-plane direction suddenly becomes antiferromagnetically aligned. The antiferromagnetic alignment of the Mn-Mn next-nearest-neighbor pairs also results in nearby Co atoms becoming nonmagnetic in a continuous manner. The transition of these Mn-Mn next-nearest-neighbor configurations to anti-parallel Mn moments results in discontinuities in the average Co and Mn moments (see FIG.~\ref{fig::atomicMag}) to approximately $1.4$ $\mu_\text{B}$ and $1.2$ $\mu_\text{B}$ respectively.

Above $c/a^\text{sub} = 1.23$, at the tetragonal distortion associated with growth on Si, the remaining next-nearest-neighbor Mn-Mn pairs undergo a ferrimagnetic transition with the weaker Mn moment aligned with the Co moments. The ferrimagnetic alignment of the Mn moments relative to each other also abruptly reduces the moment of Co atoms within their vicinity. The average Co moment steadily decreases with increasing $c/a^\text{sub}$ from $1.08$ $\mu_\text{B}$ to $0.81$ $\mu_\text{B}$, and the average Mn moment anti-parallel to the Co moments steadily increases in magnitude from $-0.05$ $\mu_\text{B}$ to $-0.27$ $\mu_\text{B}$. This is consistent with measurements of fcc Co$_{0.756}$Mn$_{0.244}$, which found a Co moment of $0.8$ $\mu_\text{B}$ and average Mn moment of $-0.1$ $\mu_\text{B}$\cite{cable1982}.

\subsubsection{Co$_{1/2}$Mn$_{1/2}$ moment}

A disordered alloy of Co$_{1/2}$Mn$_{1/2}$ was approximated by considering four ordered alloys each with two bct unit cells per supercell to allow magnetic states with two unique Mn positions, which could align antiparallel to each other when the system became antiferromagnetic in the fcc phase, which was the solution found by Wu et. al.\cite{wu2001} However, the work reported here differs from this earlier work by considering tetragonally distorted unit cells where the in-plane and out-of-plane directions are no longer equivalent, therefore two possible supercells needed to be considered.

The two supercells considered were: the $a$-class (see FIG.~\ref{fig:supercells_Co50Mn50} (a) and (c)) where the two bct unit cells were aligned along an $a$-axis and the $c$-class (see FIG.~\ref{fig:supercells_Co50Mn50} (b) and (d)) where the two bct unit cells were aligned along the $c$-axis. For each of these supercells, every possible arrangement of Co and Mn atoms was considered: one where the Mn atoms were nearest-neighbors and another where the Mn atoms were next-nearest-neighbors. The moments of these alloys were calculated as a function of $c/a^\text{sub}$ and weighted by their multiplicities to approximate a disordered alloy. See Appendix \ref{sec:superlattices}\ref{sec:Co50Mn50} for more details.

When $c/a^\text{sub} < 1.09$, the Co and Mn atoms are ferromagnetically aligned with an average Co moment ranging from $1.28$ $\mu_\text{B}$ to $1.55$ $\mu_\text{B}$ and an average Mn moment ranging from $1.55$ $\mu_\text{B}$ to $2.07$ $\mu_\text{B}$. The weakest ferromagnetic state occurs when $c/a^\text{sub} = 0.86$, which is near the hcp-like structure discussed in Section \ref{sec:theory}\ref{sec:structural} and also the tetragonal distortion associated with Co$_{1/2}$Mn$_{1/2}$ grown on MgO. However, the strongest ferromagnetic state occurs when $c/a^\text{sub} = 1.07$, which is the tetragonal distortion associated with growth on BTO substrates.

As Co$_{1/2}$Mn$_{1/2}$ undergoes a Bain deformation and $c/a^\text{sub}$ increases above $1.09$, the average Mn moment abruptly decreases to  $1.0$ $\mu_\text{B}$. This is because the regions of the disordered alloy with Mn-Mn nearest-neighbor pairs undergoes a sudden spin-reversal of one of the Mn atoms with the weaker of the two moments remaining aligned with the Co moments. The realignment of the Mn atoms causes the nearby Co moments to abruptly decrease by a small amount ($0.2$ $\mu_\text{B}$). This results in a discontinuity of the average Co and Mn moments in FIG.~\ref{fig::atomicMag}.

As $c/a^\text{sub}$ approaches $\sqrt{2}$, the fcc limit, the average Mn moment again abruptly decreases, this time to $\sim 0.1$ $\mu_\text{B}$ antiparallel to the Co moments. This is because the regions of the disordered alloy with Mn-Mn next-nearest-neighbor pairs separated by a Co atom abruptly undergo an antiferromagnetic alignment with a net-zero atomic Mn moment. The antiferromagnetic alignment of these Mn moments also results in the nearby Co atoms becoming nonmagnetic in a continuous manner. This results in a continuous suppression of the average Co moment in FIG.~\ref{fig::atomicMag}(a), but a  discontinuity in the average Mn moment in FIG.~\ref{fig::atomicMag}(b). Beyond the fcc transition, one of the ordered phases results in the Mn moments becoming aligned, while remaining antiparallel with both Co moments.

\subsubsection{Mn moment}

In this work, pure Mn is found to be ferromagnetic in the bcc phase when $a^\text{sub} < 1.1$ with an atomic moment that decreases from $1.1$ $\mu_\text{B}$ to $0.8$ $\mu_\text{B}$ with increasing $c/a^\text{sub}$ until $c/a^\text{sub} = 0.85$ where it becomes constant. This result is in good agreement with previous calculations which predicted a low-spin ferromagnetic state of $0.9$ $\mu_\text{B}$/atom for bcc Mn with a lattice constant of $2.79$ \AA\cite{fry1987,fuster1988}. These earlier calculations also predicted a high-spin state for bcc Mn which appeared beyond bcc lattice constants of $3.1$ \AA \hspace{.01in} with atomic moments in excess of $3$ $\mu_\text{B}$. While a pure Mn bcc system with lattice constants this large have not been experimentally attained, the theoretical high-spin state is created by the same mechanism that results in large Mn moments when alloyed with other ferromagnetic materials, thus leading to larger spacings between Mn atoms.

When $c/a^\text{sub} > 1.1$, Mn becomes antiferromagnetic with a sublattice moment of $\pm 0.8$ $\mu_\text{B}$. As $c/a^\text{sub}$ increases further and the lattice transitions to fct, the sublattice moment increases to $1.6$ $\mu_\text{B}$. By the time $c/a^\text{sub} = \sqrt{2}$, the sublattice moment has fallen once again down to $1.3$ $\mu_\text{B}$. Previous calculations\cite{wu2001} of fcc Mn also predicted antiferromagnetic alignment between the Mn moments, albeit with a larger sublattice moment of $\pm2.1$ $\mu_\text{B}$.

\section{\label{sec:conclusion}Conclusion}

The composition-dependence of the average magnetic moment of bct Co$_{1-x}$Mn$_x$ films on MgO(001) is in good agreement with the Slater-Pauling curve\cite{slater1936,pauling1938,williams1983}, especially for low Mn concentrations ($x < 1/4$). In fact, these films offer a competitive average atomic moment to the pinnacle of the Slater-Pauling curve: Fe$_{0.65}$Co$_{0.35}$ ($n_e/\text{atom} = 26.35$) which has a moment of $2.42$ $\mu_\text{B}/\text{atom}$, where Co$_{0.76}$Mn$_{0.24}$ ($n_e/\text{atom} = 26.52$) has a moment of $(2.52 \pm 0.07)$ $\mu_\text{B}/\text{atom}$. 

The calculations performed in this work reveal that the $c/a^\text{sub}$-ratio associated with Co$_{3/4}$Mn$_{1/4}$ grown on MgO(001) substrates both offered the weakest ferromagnetic state (see FIG.~\ref{fig::atomicMag}) and is near a bcc $\rightarrow$ hcp transition, observed by Wang, et. al.\cite{wang2021} Growth of forced Co$_{1-x}$Mn$_x$ films on substrates with slightly lower lattice constants such as GaAs(001) or BaTiO$_3$(110) could lead to further enhanced Mn atomic moments by $\sim0.20$ - $0.35$ $\mu_\text{B}$, resulting in an increased average moment of $\sim0.05$ - $0.09$ $\mu_\text{B}$ respectively for $x = 1/4$. Growth on these substrates could also result in a higher structural stability of the bct phase of Co$_{1-x}$Mn$_x$, thus allowing for the growth of thicker films than are attainable on MgO(001) substrates.

BaTiO$_3$ barriers are a feasible replacement for the MgO barriers typically used in MTJs for Co-based materials\cite{cao2011}. However, BaTiO$_3$ is ferroelectric and will result in a coexisting tunneling electroresistance (TER) and TMR effect. In principle, the coexistence of these effects would lead to a large tunneling electromagnetoresistance (TEMR) effect\cite{caffrey2012}. However, it would require not only controlling the moment of the free layer, but the polarization of the ferroelectric barrier as well, potentially limiting their practical usefulness.

While GaAs can be used as a tunneling barrier for MTJs\cite{moser2007}, its use severely limits the TMR effect and it can be difficult to prevent As interdiffusion into the Co$_{1-x}$Mn$_x$ films. At the moment, MgO-based tunnel junctions are effectively the only option for achieving a large TMR effect. However, it is possible to grow MTJ stacks on other substrates, such as GaAs mediated by a MgO layer\cite{bowen2001} to prevent As poisoning of the bottom layer of the MTJ. In order to preserve the in-plane lattice constant the GaAs substrate imposes upon Co$_{1-x}$Mn$_x$ films, a thin MgO buffer layer could be used. It has been demonstrated that thin MgO layers can change their lattice constant commensurate with the material they're grown on\cite{elphick2021}, and shouldn't change the imposed in-plane lattice constant from the GaAs substrate on subsequent layers. This should allow for an enhancement of the magnetic moments throughout the MTJ. GaAs has a long history of use as a substrate (the first bcc Co sample was deposited on GaAs\cite{prinz1985}) and atomically flat high-quality materials are readily-available.

\section*{Acknowledgements}

This work was conducted at the NSF MonArk Quantum Foundry which is supported from the NSF Q-AMASE-I program Award No. DMR1906383. The authors would like to thank Frank A Schooner III for many useful conversations.

\appendix

\section{\label{sec:superlattices} Superlattice Calculations and Averages}

\begin{figure*}[t]
\includegraphics[width=16cm]{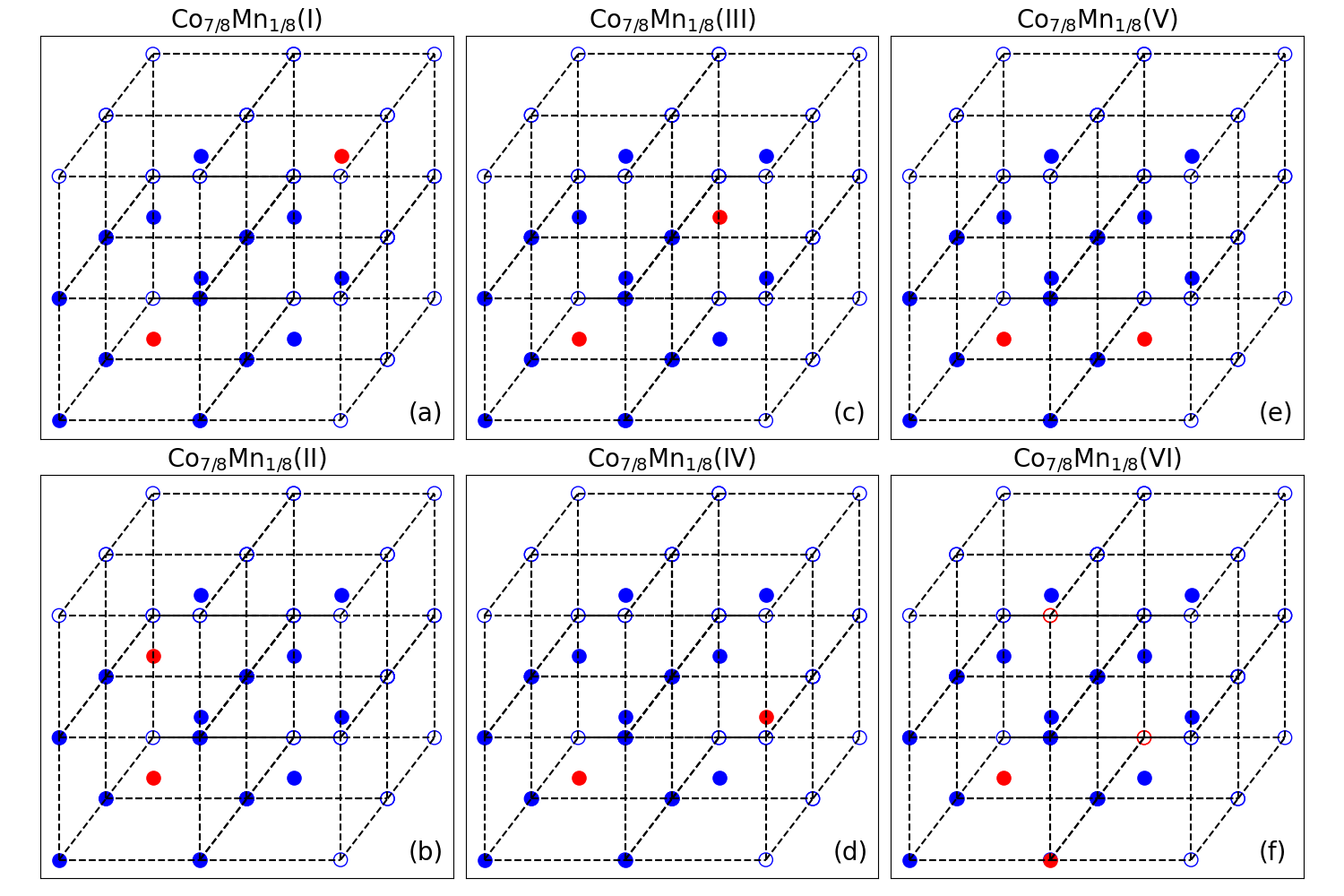}
%\caption{Schematic diagram of bct supercells used to model various orderings of Co$_{1-x}$Mn$_x$ alloys. Blue lattice sites represent those which are actually within the show unit cells (outlined by the black dashed lines), whereas red lattice sites belong to other repeated cells and merely included in this figure for aesthetic purposes. (a) Supercell consisting of 2 bct unit cells with 4 atoms, stacked along the $c$-axis. (b) Supercell consisting of 2 bct unit cells with 4 atoms, stacked along an $a$-axis. (c) Supercell consisting of 4 bct unit cells with 8 atoms, stacked along an $a$- and $c$-axis. (d) Supercell consisting of 4 bct unit cells with 8 atoms, stacked along both $a$-axes. (e) Supercell consisting of 8 bct unit cells with 16 atoms, forming cube of cells.}
\caption{Supercell configurations of ordered Co$_{7/8}$Mn$_{1/8}$ consisting of 8 bct cells arranged in a cube where blue and red solid circles indicate Co and Mn atoms within the supercell, hollow circles indicate atoms in adjacent repeated cells. (a) The (I) ordered alloy has Mn atoms at the centers of bct cells at opposite corners of the supercell. (b) The (II) ordered alloy has Mn atoms at the centers of nearest-neighbor cells along the $c$-axis. (c) The (III) ordered alloy has Mn atoms at the centers of next-nearest-neighbor cells $\sqrt{a^2+c^2}$ away. (d) The (IV) ordered alloy has Mn atoms at the centers of next-nearest-neighbor cells $a\sqrt{2}$ away. (e) The (V) ordered alloy has Mn atoms at the centers of nearest-neighbor cells along an $a$-axis. (f) The (VI) ordered alloy has a nearest-neighbor Mn pair.}
\label{fig:supercells_Co875Mn125}
\end{figure*}

\subsection{\label{sec:Co875Mn125} Co$_{7/8}$Mn$_{1/8}$}

\begin{table}[h!]
\caption{\label{tab:Co875Mn125_orders} Distance between nearest-neighbor Mn atoms in supercell ($d_\text{Mn-Mn}$), and multiplicity ($\Omega$) of Co and Mn atoms for the 6 unique configurations of Co$_{7/8}$Mn$_{1/8}$ using the supercells shown in FIG.~\ref{fig:supercells_Co875Mn125}.}
\begin{ruledtabular}
\begin{tabular}{c|cccc}
 & $d_\text{Mn-Mn}$ & $\Omega$ & $\Delta E$ & $\mu^\text{Mn}_{\text{MgO}}$\\
\hline
Co$_{7/8}$Mn$_{1/8}$(I) & $a\sqrt{2 + \frac{c^2}{a^2}}$ & $8$ & $2.97$ eV & $2.64$ $\mu_\text{B}$\\
Co$_{7/8}$Mn$_{1/8}$(II) & $c$ & $8$ & $5.09$ eV & $2.38$ $\mu_\text{B}$\\
Co$_{7/8}$Mn$_{1/8}$(III) & $a\sqrt{1 + \frac{c^2}{a^2}}$ & $16$ & $11.4$ eV &$2.55$ $\mu_\text{B}$ \\
Co$_{7/8}$Mn$_{1/8}$(IV) & $a\sqrt{2}$ & $8$ & $0$ eV & $2.24$ $\mu_\text{B}$\\\
Co$_{7/8}$Mn$_{1/8}$(V) & $a$ & $16$ & $6.80$ eV & $2.43$ $\mu_\text{B}$\\
Co$_{7/8}$Mn$_{1/8}$(VI) & $a\sqrt{\frac{1}{2} + \frac{1}{4}\frac{c^2}{a^2}}$ & $64$ & $33.9$ eV & $-1.94$ $\mu_\text{B}$\\
\end{tabular}
\end{ruledtabular}
\end{table}

The Co$_{7/8}$Mn$_{1/8}$ calculation was performed to add an additional calibration point for the Mn atomic moment data, as such it was only calculated for $a^\text{sub} = 2.91$\AA \hspace{.01in} with the calculated $c/a^\text{sub} = 0.86$ ratio which relaxed the lattice for that in-plane lattice constant. 

The Co$_{7/8}$Mn$_{1/8}$ supercell consisted of 8 typical bct unit cells forming a $2 \times 2 \times 2$ cube of bct units cells (see FIG.~\ref{fig:supercells_Co875Mn125}) consisting of 2 Mn atoms and 14 Co atoms. The 2 Mn atoms were arranged in every possible unique combination of the 16 sites in the supercell. This resulted in six unique locations for the two Mn atoms, as can be seen in TABLE~\ref{tab:Co875Mn125_orders}. 

Every phase of Co$_{7/8}$Mn$_{1/8}$, except for the (VI) phase which had a pair of nearest-neighbor Mn atoms in its supercell, resulted in a ferromagnetic alignment of the Mn moments with the Co moments and a magnitude ranging from $2.2$ - $2.6$ $\mu_\text{B}$. The Co$_{7/8}$Mn$_{1/8}$(I) phase had the largest distance between Mn atoms, and unsurprisingly resulted in the largest Mn moment of $2.64$ $\mu_\text{B}$. The (VI) phase resulted in the Mn moments becoming anti-aligned with the Co moments with a magnitude of $1.94$ $\mu_\text{B}$. Because of this anti-alignment with the Co moments, this phase was excluded from the weighted average of these moments that was used as a calibration point for the experimental Mn moments in FIG.~\ref{fig:atomicMag_exp}(a).

\subsection{\label{sec:Co75Mn25} Co$_{3/4}$Mn$_{1/4}$}

\begin{figure*}[t]
\includegraphics[width=17.5cm]{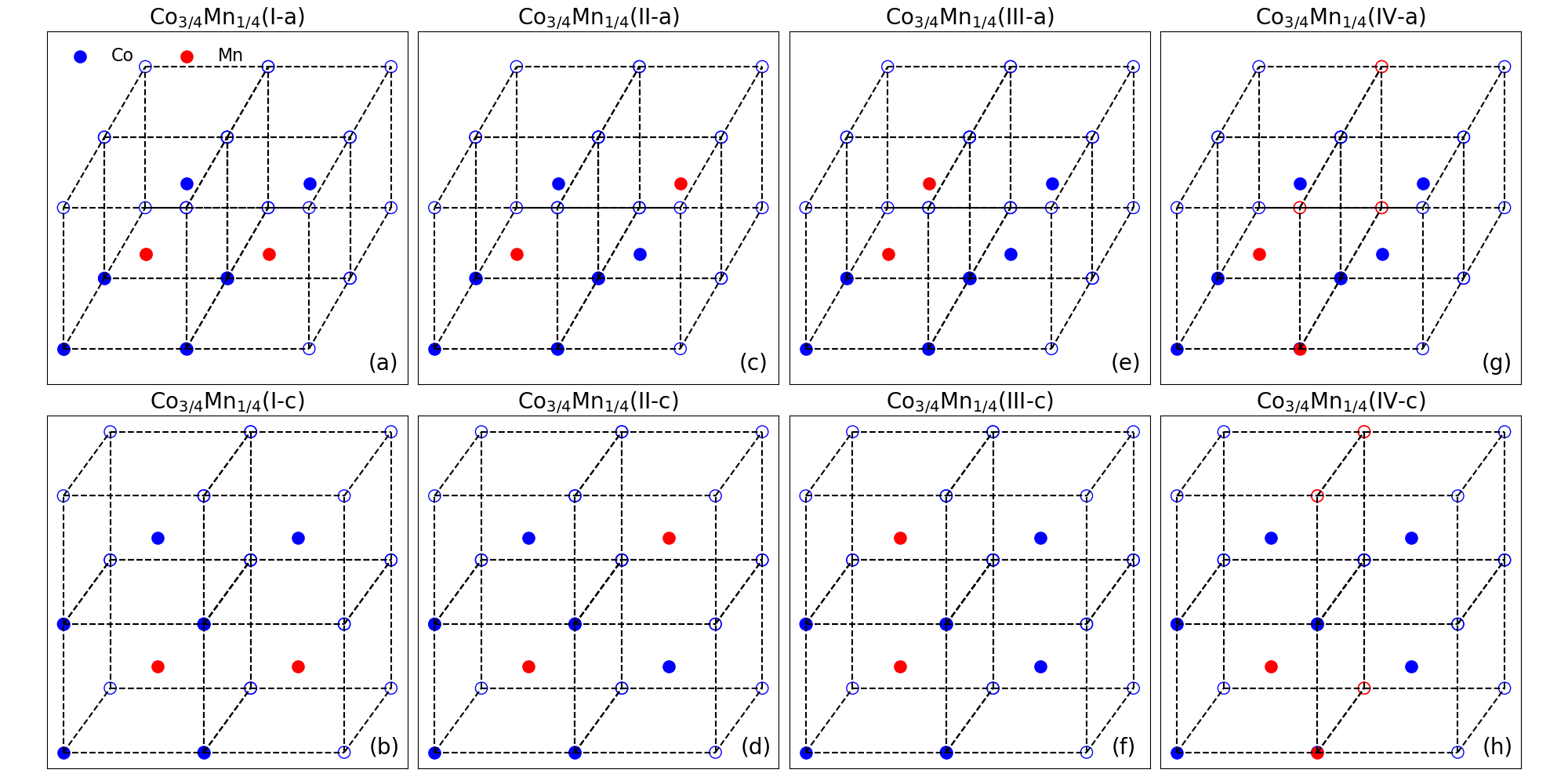}
\caption{Supercell configurations of ordered Co$_{3/4}$Mn$_{1/4}$ consisting of 4 bct cells where blue and red solid circles indicate Co and Mn atoms within the supercell, hollow circles indicate atoms in adjacent repeated cells. The $a$-class of supercells has 4 bct unit cells arranged in a square with its axes both aligned along the $a$-axes. The $c$-class of supercells has 4 bct unit cells arranged in a square with one axis aligned along an $a$-axis and the other along the $c$-axis. (a) The (I-a) ordered alloy has Mn atoms at the centers of nearest-neighbor cells along an $a$-axis. (b) The (I-c) ordered alloy has Mn atoms at the centers of nearest-neighbor cells along an $a$-axis. (c) The (II-a) ordered alloy has Mn atoms at the centers of next-nearest-neighbor cells. (d) The (II-c) ordered alloy has Mn atoms at the centers of next-nearest-neighbor cells. (e) The (III-a) ordered alloy is identical to the (I-a) alloy and therefore wasn't calculated separately, but rather doubled the multiplicity of (I-a). (f) The (III-c) ordered alloy has Mn atoms at the centers of nearest-neighbor cells along the $c$-axis. (g) The (IV-a) ordered alloy with one Mn atom at the center of a bct cell and at the corner of another, resulting in a nearest-neighbor Mn pair. (h) The (IV-c) ordered alloy with one Mn atom at the center of a bct cell and at the corner of another, resulting in a nearest-neighbor Mn pair.}
\label{fig:supercells_Co75Mn25}
\end{figure*}

In order to model the Co$_{3/4}$Mn$_{1/4}$ alloy with 2 Mn atoms in the supercell, 4 bct unit cells were considered. Co$_{3/4}$Mn$_{1/4}$(I-c), (II-c), (III-c), and (IV-c) consisting of a square of 4 bct unit cells with one side lying along an $a$-axis and the other along the $c$-axis. The $\mathbf{k}$-grid associated with these $x = 1/4$ phases was reduced to be consistent with the symmetry of the supercell and was a $10 \times 5 \times 5$ grid. Co$_{1/2}$Mn$_{1/2}$(I-a), (II-a), and (IV-a) consisted of a square of 4 bct unit cells lying in both $a$-axes. The $\mathbf{k}$-grid associated with these $x = 1/4$ phases was reduced to be consistent with the symmetry of the supercell and was a $5 \times 5 \times 10$ grid. The reason for the lack of a (III-a) phase is that it's identical to the (I-a) phase, and is why the multiplicity of (I-a) is double that of (I-c).

\begin{figure}[t!]
\includegraphics[width=8.5cm]{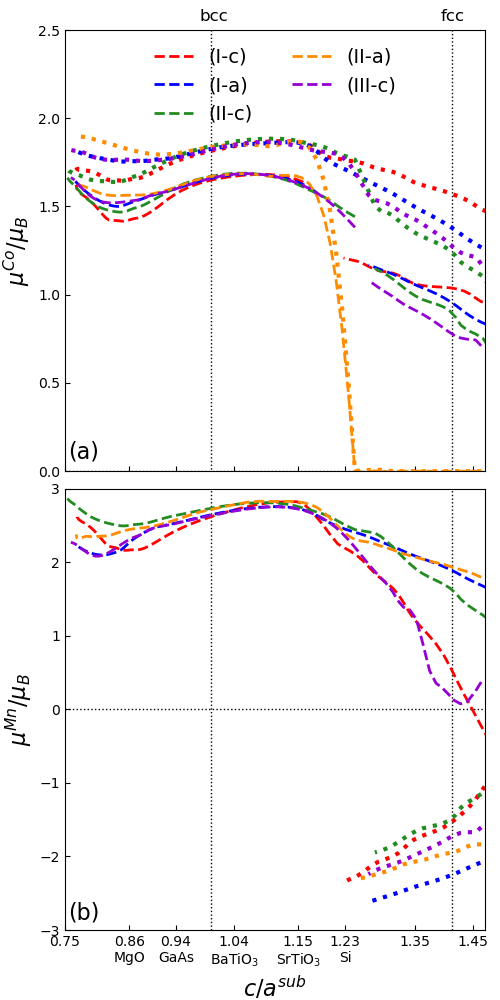}
\caption{The tick-marks for $c/a^\text{sub}$ are labeled to represent the corresponding $c/a^\text{sub}$-ratio from common experimental substrates on Co$_{3/4}$Mn$_{1/4}$.}
\label{fig:atomicMag_Co75Mn25}
\end{figure}

The (I-a) and (III-c) phases are structurally identical with sheets of pure bct Co separated by sheets of bct Co$_{1/2}$Mn$_{1/2}$ both of which are aligned along an $a$-axis and a $c$-axis. This results in nearly identical ferromagnetic atomic moments (see the blue and purple curves in FIG.~\ref{fig:atomicMag_Co75Mn25}). The (I-c) phase is similar, with sheets of pure bct Co separated by layers of bct Co$_{1/2}$Mn$_{1/2}$ which are aligned along the $a$-axes, resulting in lower $d_\text{Mn-Mn}$-values when $c/a^\text{sub} < 1$, lower Mn atomic moments, higher $d_\text{Mn-Mn}$-values when $c/a^\text{sub} > 1$, and higher Mn atomic moments. 

In general, ferromagnetic Mn moments become stronger as their distance increases. As these materials undergo a Bain deformation from bcc $\rightarrow$ fcc, the Mn atoms become ferrimagnetic with larger Mn moments anti-aligned with the ferromagnetic Co moments. When this occurs, the (I-a) and (III-c) solutions are structurally identical but not magnetically identical due to the Mn moments being anti-aligned to their nearest-neighbors along the $a$-axis, but aligned along the $c$-axis for (I-a), and the opposite is true for (III-c). This difference occurs from the choice of supercell effectively acting as the magnetic cell. The ferrimagnetic Mn solution of (I-c) results in anti-aligned Mn moments along one $a$-axis and aligned Mn moments along the other $a$-axis.

The (II-a) and (II-c) phases are only structurally identical for bcc lattices, however they do both result in the largest $d_\text{Mn-Mn}$-values of the considered ordered alloys in this work. These ordered alloys have the largest Mn moments and are the most energetically favorable phases of Co$_{3/4}$Mn$_{1/4}$. The (II-a) phase results in columns of Co$_{1/2}$Mn$_{1/2}$ bct cells along the $c$-axis, while the (II-c) phase similarly results in columns of Co$_{1/2}$Mn$_{1/2}$ bct cells along an $a$-axis all surrounded by pure bct Co cells. Interestingly, the (II-c) phase has the highest Mn moment when $c/a^\text{sub} < 1$ despite the nearby Mn atoms within the supercell being closer than those of the (II-a) phase because the Mn atom located within the next bct Co$_{1/2}$Mn$_{1/2}$ in the column of (II-c) is located one in-plane lattce constant ($a$) away, whereas (II-a) is located one out-of-plane lattice constant ($c$) away. This same effect also results in the Mn moments of (II-c) being slightly lower than those of (II-a) when $c/a^\text{sub} > 1$. As these phases go from bcc $\rightarrow$ fcc the Mn moments become anti-aligned. The (II-a) phase transitions from a ferromagnetic Mn to an antiferromagnetic one with zero net Co moment. The ferromagnetic $\rightarrow$ nonmagnetic transition in the Co atomic moments of the (II-a) phase is continuous, unlike the transitions to lower-moment ferromagnetic states present in the other ordered alloys (see FIG.~\ref{fig:atomicMag_Co75Mn25}(a)). The (II-c) phase transitions from a ferromagnetic Mn to an ferrimagnetic one with a stronger Mn moment anti-aligned with the ferromagnetic Co moments. 

The (IV-a) and (IV-c) phases are ferrimagnetic over the entire $c/a^\text{sub}$-range investigated in this work. Additionally, they are energetically unfavorable states by $\sim15$ eV. Because of this, they were excluded from the weighted average for the disordered alloy.

\begin{table}[h!]
\caption{\label{tab:Co75Mn25_orders} Distance between nearest-neighbor Mn atoms ($d_\text{Mn-Mn}$), and multiplicity ($\Omega$) of Co and Mn atoms for the 7 unique configurations of Co$_{3/4}$Mn$_{1/4}$ using the supercells shown in FIG.~\ref{fig:supercells_Co75Mn25}.}
\begin{ruledtabular}
\begin{tabular}{c|cccc}
 & $d_\text{Mn-Mn}$ & $\Omega$ & $\Delta E$ & $\mu^\text{Mn}_{\text{MgO}}$\\
\hline
Co$_{3/4}$Mn$_{1/4}$(I-c) & $a$ & $4$ & $5.85$ eV & $2.17$ $\mu_\text{B}$\\
Co$_{3/4}$Mn$_{1/4}$(I-a) & $a$ & $8$ & $7.27$ eV & $2.29$ $\mu_\text{B}$\\
Co$_{3/4}$Mn$_{1/4}$(II-c) & $a\sqrt{1+\frac{c^2}{a^2}}$ & $4$ & $1.61$ eV & $2.50$ $\mu_\text{B}$\\
Co$_{3/4}$Mn$_{1/4}$(II-a) & $a\sqrt{2}$ & $4$ & $0$ eV & $2.46$ $\mu_\text{B}$\\
Co$_{3/4}$Mn$_{1/4}$(III-c) & $c$ & $4$ & $10.5$ eV & $2.26$ $\mu_\text{B}$\\
Co$_{3/4}$Mn$_{1/4}$(IV-c) & $a\sqrt{\frac{1}{2} + \frac{1}{4}\frac{c^2}{a^2}}$ & $16$ & $26.2$ eV & $1.99$ $\mu_\text{B}$\\
 & & & & $-2.16$ $\mu_\text{B}$\\
Co$_{3/4}$Mn$_{1/4}$(IV-a) & $a\sqrt{\frac{1}{2} + \frac{1}{4}\frac{c^2}{a^2}}$ & $16$ & $12.3$ eV & $2.04$ $\mu_\text{B}$\\
 & & & & $-1.80$ $\mu_\text{B}$\\
\end{tabular}
\end{ruledtabular}
\end{table} 

\subsection{\label{sec:Co50Mn50} Co$_{1/2}$Mn$_{1/2}$}

\begin{figure}[t!]
\begin{centering}
\includegraphics[width=8cm]{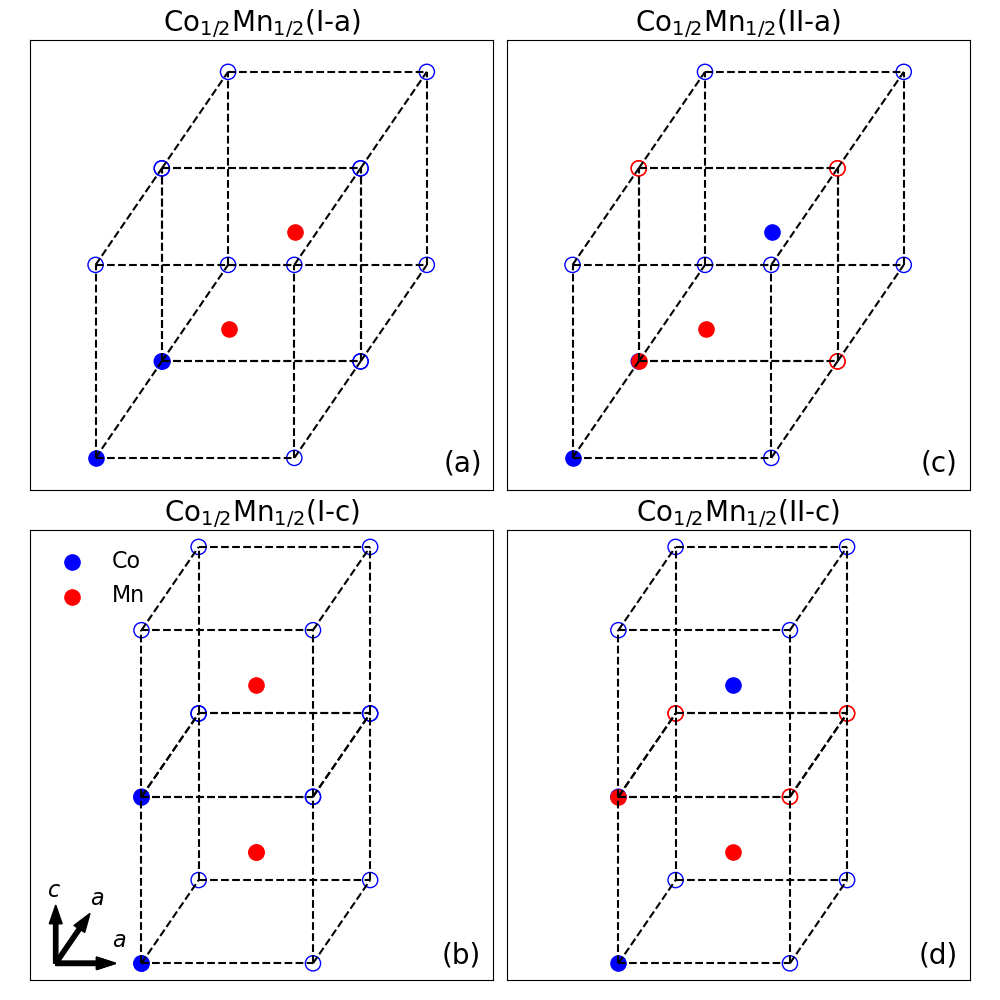}
\caption{Supercell configurations of ordered Co$_{1/2}$Mn$_{1/2}$ consisting of 2 bct cells where blue and red solid circles indicate Co and Mn atoms within the supercell, hollow circles indicate atoms in adjacent repeated cells. (a) Supercell of (I-a) ordered alloy with 2 bct cells stacked along an $a$-axis with Mn atoms at the center of the bct cells. (b) Supercell of (I-c) ordered alloy with 2 bct cells stacked along the $c$-axis with Mn atoms at the center of the bct cells. (c) Supercell of (II-a) ordered alloy with 2 bct cells stacked along an $a$-axis with Mn atoms at the center of one bct cell and at the corner of the other. (d) Supercell of (II-a) ordered alloy with 2 bct cells stacked along the $c$-axis with Mn atoms at the center of one bct cell and at the corner of the other.} 
\label{fig:supercells_Co50Mn50}
\end{centering}
\end{figure}

In order to model the Co$_{1/2}$Mn$_{1/2}$ alloy with 2 Mn atoms in the supercell, 2 bct unit cells were considered. Co$_{1/2}$Mn$_{1/2}$(I-c) and (II-c) consisted of 2 bct unit cells stacked along the unique $c$-axis. The $\mathbf{k}$-grid associated with these $x = 1/2$ phases was reduced to be consistent with the symmetry of the supercell and was a $10 \times 10 \times 5$ grid. Co$_{1/2}$Mn$_{1/2}$(I-a) and (II-a) consisted of 2 bct unit cells stacked along one of the $a$-axes. The $\mathbf{k}$-grid associated with these $x = 1/2$ phases was reduced to be consistent with the symmetry of the supercell and was a $10 \times 5 \times 10$ grid.

\begin{table}[h!]
\caption{\label{tab:Co50Mn50_orders} Distance between nearest-neighbor Mn atoms ($d_\text{Mn-Mn}$), and multiplicity ($\Omega$) of Co and Mn atoms for the 4 unique configurations of Co$_{1/2}$Mn$_{1/2}$ using the supercells shown in FIG.~\ref{fig:supercells_Co50Mn50}.}
\begin{ruledtabular}
\begin{tabular}{c|cccc}
 & $d_\text{Mn-Mn}$ & $\Omega$ & $\Delta E$ & $\mu^\text{Mn}_{\text{MgO}}$ \\
\hline
Co$_{1/2}$Mn$_{1/2}$(I-c) & $c$ & $2$ & $0$ eV & $2.30$ $\mu_\text{B}$\\
Co$_{1/2}$Mn$_{1/2}$(I-a) & $a$ & $2$ & $0.68$ eV & $2.30$ $\mu_\text{B}$\\
Co$_{1/2}$Mn$_{1/2}$(II-c) & $a\sqrt{\frac{1}{2} + \frac{1}{4}\frac{c^2}{a^2}}$ & $4$ & $8.03$ eV & $1.02$ $\mu_\text{B}$\\
Co$_{1/2}$Mn$_{1/2}$(II-a) & $a\sqrt{\frac{1}{2} + \frac{1}{4}\frac{c^2}{a^2}}$ & $4$ & $2.93$ eV & $1.36$ $\mu_\text{B}$\\
\end{tabular}
\end{ruledtabular}
\end{table}

\begin{figure}[t!]
\includegraphics[width=8.5cm]{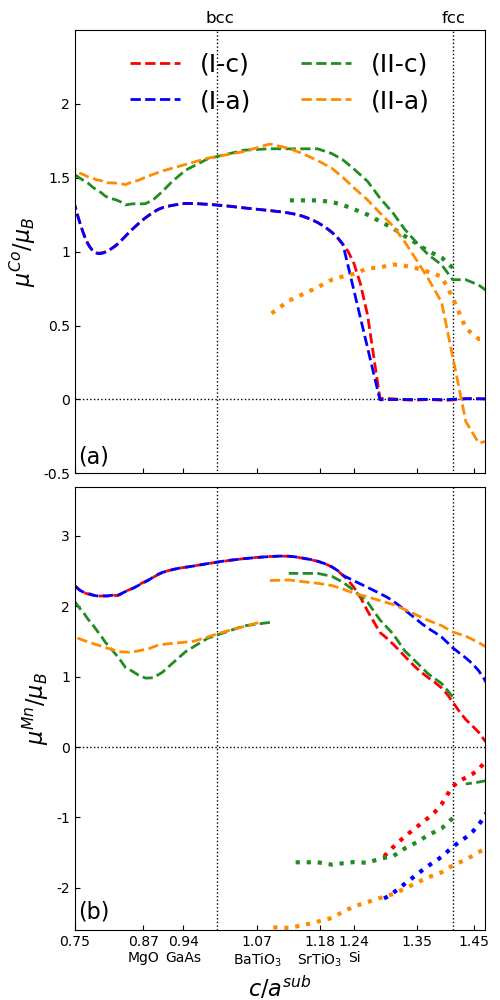}
\caption{The tick-marks for $c/a^\text{sub}$ are labeled to represent the corresponding $c/a^\text{sub}$-ratio from common experimental substrates on Co$_{1/2}$Mn$_{1/2}$.}
\label{fig:atomicMag_Co50Mn50}
\end{figure}

Co$_{1/2}$Mn$_{1/2}$(I-c) and Co$_{1/2}$Mn$_{1/2}$(I-a) are structurally identical systems consisting of bct cells with Co atoms at their vertices and Mn atoms at their centers, which is the furthest apart single Mn atoms can be from each other, either $c$ or $a$ apart, at this composition. Because these are structurally identical, the ferromagnetic solutions for these are identical solutions, see FIG.~\ref{fig:atomicMag_Co50Mn50}. The ferromagnetic solution of the (I-c) and (I-a) phases results in a stronger Mn moment, but a weaker Co moment than the (II-c) and (II-a) phases. This is due to the fact that Mn moments become stronger as the Mn atoms are separated, whereas Co moments become weaker as the Co atoms are separated. As the (I-c) and (I-a) phases undergo a Bain deformation from bcc $\rightarrow$ fcc, the systems gradually transition from being ferromagnetic to the Mn atoms aligning antiferromagnetically and the Co atoms becoming nonmagnetic, these results are consistent with previous calculations\cite{wu2001} of a similar ordered Co$_{1/2}$Mn$_{1/2}$ alloy. The difference between the antiferromagnetic solutions of the (I-c) and (I-a) phases is that the Mn moments are only antiparallel to their nearest-neighbors along a different single axis and parallel to them along the other two axes. The (I-c) phase has Mn atoms which which are aligned antiparallel along the $c$-axis and parallel along the $a$-axes. Whereas the (I-a) phase has Mn atoms that are antiparallel along one of the $a$-axes and parallel along the other $a$-axis and $c$-axis. The antiferromagnetic (I-c) phase is weaker because $c/a > 1$, so the anti-aligned Mn moments which are separated by a distance of $c$ are further away than the aligned Mn moments which are separated by a distance of $a$. In the antiferromagnetic (I-a) phase, the antiparallel Mn moments are only separated by $a$ and some parallel moments are separated by $a$ while the other half are separated by $c$.

The (II-c) and (II-a) phases aren't identical structurally, except for in the bcc phase. This results in different ferromagnetic solutions away from $c/a = 1$. The structural difference between these phases is that (II-c) has planes of bct Mn cells separated by planes of bct Co cells, which lie along the $a$-axes. Whereas, these planes in the (II-a) phase are aligned along the $c$-axis and an $a$-axis. The Mn atoms in the (II-c) and (II-a) phases are significantly closer than the (I-c) and (I-a) phases, because of this the Mn ferromagnetic moments are significantly reduced and the Co moments are enhanced. The ferromagnetic states of the (II-c) and (II-a) phases are less stable against a Bain deformation than the (I-c) and (I-a) phases. Additionally, the (II-c) and (II-a) phases also have a sharp transition from a ferromagnetic solution to a ferrimagnetic solution where the Mn moments abruptly become antiferromagnetically aligned with equal magnitudes but antiparallel directions and the Co moments remain aligned to one of the Mn moments but have different magnitudes. Beyond the transition to an fcc lattice structure, the Mn moments of the (II-c) phase realign with each other with equal magnitudes and are anti-aligned with the Co moments which also once again have the same magnitude. The Co moments of the (II-a) phase become anti-aligned beyond the transition to an fcc lattice, similar to the anti-alignment in the Mn moments.

The evolution of these magnetic moments as a function of the Bain deformation can be averaged, weighted by the multiplicity of the phases (see TABLE~\ref{tab:Co50Mn50_orders}), to approximate a disordered alloy. 

\section{\label{sec:NiMn} fcc Ni$_{1-x}$Mn$_{x}$ atomic moments}

In order to compare the downturn of the average fcc Ni$_{1-x}$Mn$_x$ moment to that of the average bct Co$_{1-x}$Mn$_x$ moment, the Ni and Mn atomic moments for fcc Ni$_{7/8}$Mn$_{1/8}$ and Ni$_{3/4}$Mn$_{1/4}$ were calculated using the same method as the Co and Mn moments for Co$_{1-x}$Mn$_x$ at the same Mn concentration outlined in Appendix~\ref{sec:superlattices}\ref{sec:Co875Mn125} and \ref{sec:superlattices}\ref{sec:Co75Mn25}. First, the fcc lattice constants were found by setting $c = a\sqrt{2}$ in the superlattices of Ni$_{3/4}$Mn$_{1/4}$(II-c) and Ni$_{7/8}$Mn$_{1/8}$(III) and minimizing the total energy relative to $a$, similar to how the bcc lattice constants were solved for previously. The fcc lattice constant of Ni$_{7/8}$Mn$_{1/8}$ was found to be $3.52$\AA, and the fcc lattice constant of Ni$_{3/4}$Mn$_{1/4}$ was found to be $3.51$\AA. 

\begin{table}
\caption{\label{tab:NiMn} Average atomic moments for ordered fcc Ni$_{1-x}$Mn$_x$ alloys when $x = 1/8$ and $1/4$.}
\begin{ruledtabular}
\begin{tabular}{c|ccc}
  & $d_\text{Mn-Mn}$ & $\mu^\text{Mn}$ & $\mu^\text{Ni}$\\
\hline
Ni$_{7/8}$Mn$_{1/8}$(I) & $a^\text{fcc}\sqrt{2}$ & $3.11$ $\mu_\text{B}$ & $0.64$ $\mu_\text{B}$\\
Ni$_{7/8}$Mn$_{1/8}$(II) & $a^\text{fcc}$ & $3.03$ $\mu_\text{B}$ & $0.60$ $\mu_\text{B}$\\
Ni$_{7/8}$Mn$_{1/8}$(III) & $a^\text{fcc} \sqrt{3/2}$ & $3.00$ $\mu_\text{B}$ & $0.61$ $\mu_\text{B}$\\
Ni$_{7/8}$Mn$_{1/8}$(IV) & $a^\text{fcc}$ & $3.01$ $\mu_\text{B}$ & $0.60$ $\mu_\text{B}$\\
Ni$_{7/8}$Mn$_{1/8}$(V) & $a^\text{fcc}/\sqrt{2}$ & $2.78$ $\mu_\text{B}$ & $0.61$ $\mu_\text{B}$\\
Ni$_{7/8}$Mn$_{1/8}$(VI) & $a^\text{fcc}/\sqrt{2}$ & $2.98$ $\mu_\text{B}$ & $0.62$ $\mu_\text{B}$\\
\hline
Ni$_{3/4}$Mn$_{1/4}$(I-c) & $a^\text{fcc}/\sqrt{2}$ & $1.68$ $\mu_\text{B}$ & $0.35$ $\mu_\text{B}$\\
Ni$_{3/4}$Mn$_{1/4}$(I-a) & $a^\text{fcc}/\sqrt{2}$ & 2.52 $\mu_\text{B}$ & 0.41 $\mu_\text{B}$\\
Ni$_{3/4}$Mn$_{1/4}$(II-c) & $a^\text{fcc}\sqrt{3/2}$ & $2.77$ $\mu_\text{B}$ & $0.63$ $\mu_\text{B}$\\
Ni$_{3/4}$Mn$_{1/4}$(II-a) & $a^\text{fcc}$ & $2.99$ $\mu_\text{B}$ & $0.49$ $\mu_\text{B}$\\
Ni$_{3/4}$Mn$_{1/4}$(III-c) & $a^\text{fcc}$ & $2.52$ $\mu_\text{B}$ & $0.41$ $\mu_\text{B}$\\
Ni$_{3/4}$Mn$_{1/4}$(IV-c) & $a^\text{fcc}$ & $1.98$ $\mu_\text{B}$ & $0.46$ $\mu_\text{B}$\\
Ni$_{3/4}$Mn$_{1/4}$(IV-a) & $a^\text{fcc}$ & $2.81$ $\mu_\text{B}$ & $0.55$ $\mu_\text{B}$\\
\end{tabular}
\end{ruledtabular}
\end{table}

Atomic moment calculations of Ni$_{7/8}$Mn$_{1/8}$ and Ni$_{3/4}$Mn$_{1/4}$ were performed identically to Co$_{1-x}$Mn$_x$ alloys of the same Mn concentration, with the only exception being that just the fcc limit was calculated. At low Mn concentrations of $x = 1/8$, the Mn atomic moment is $\sim 3$ $\mu_\text{B}$ and the Ni atomic moment is $\sim 0.6$ $\mu_\text{B}$ for every fcc Ni$_{1-x}$Mn$_x$ ordering. This calculation is consistent with measurements done by Cable et. al.\cite{cable1974} which observed average Mn and Ni atomic moments of $(3.18 \pm 0.07)$ $\mu_\text{B}$ and $0.525$ respectively. As the Mn concentration was increased to $x = 1/4$, some Mn atomic moments retained a high-spin with moments ranging from $2.5$ - $3.0$ $\mu_\text{B}$. This calculation is consistent with measurements\cite{shull1955} on ordered fcc Ni$_{3/4}$Mn$_{1/4}$ which found an average atomic Mn and Ni moment of $(3.18 \pm 0.25)$ $\mu_\text{B}$ and $(0.30 \pm 0.05)$ $\mu_\text{B}$ respectively. The existence of a high-spin ordered fcc Ni$_{3/4}$Mn$_{1/4}$ state is also consistent with previous calculations\cite{jing2009,nautiyal1993}, which found Mn atomic moments ranging from $2.7$ - $3.2$ $\mu_\text{B}$ and Ni moments ranging from $0.4$ - $0.32$ $\mu_\text{B}$. However, some orderings of Ni$_{3/4}$Mn$_{1/4}$ calculated in this work were found to occupy a low-spin Mn state with a moment ranging from $1.68$ - $2.99$ $\mu_\text{B}$. The Mn moment of atoms in Ni$_{3/4}$Mn$_{1/4}$ alloys depends not only on the distance between Mn-Mn pairs, but proximity to other Mn atoms as well. This calculation is consistent with the reduction in the average atomic moment of Ni$_{1-x}$Mn$_x$ observed by Cable et. al.\cite{cable1974} who observed both decreasing Mn and Ni moments with increasing Mn concentration above $x = 1/8$.

\newpage

\bibliography{ref}

\end{document}